\documentclass[pss]{wiley2sp} % provides pss two-column style
\usepackage{amsmath}
\usepackage{bm}             

\usepackage{multirow}
\usepackage{mathtools}
\usepackage{amssymb}
\begin{document}

% Title of the article
\title{Ion Impact Induced Ultrafast Electron Dynamics in Correlated Materials and Finite Graphene Clusters}

% Authors
\author{%
Michael Bonitz\textsuperscript{\Ast,\textsf{\bfseries 1}}, 
  Karsten Balzer\textsuperscript{\textsf{\bfseries 2}},
  Niclas Schl\"unzen\textsuperscript{\textsf{\bfseries 1}},
  Maximilian Rodriguez Rasmussen\textsuperscript{\textsf{\bfseries 1}}  and
  Jan-Philip Joost\textsuperscript{\textsf{\bfseries 1}}
  }

%E-mail-address of corresponding author
\mail{e-mail
  \textsf{bonitz@theo-physik.uni-kiel.de}, Phone: +49 431 880 4122}

% author's affiliations/addresses
\institute{%
  \textsuperscript{1}\,Institut f\"ur Theoretische Physik und Astrophysik, Christian-Albrechts-Universit\"at zu Kiel, 24098 Kiel, Germany,\\
   \textsuperscript{2}\,Rechenzentrum, Christian-Albrechts-Universit\"at zu Kiel, 24098 Kiel, Germany}

% Please select about four verbal keywords for your manuscript.

\abstract{Strongly correlated systems of fermions have an interesting phase diagram arising from the Hubbard gap. Excitation across the gap leads to the formation of doubly occupied lattice sites (doublons) which offers interesting electronic and optical properties.
Moreover, when the system is driven out of equilibrium interesting collective dynamics may arise that are related to the spatial propagation of doublons. 
%
%In the past a variety of methods have been proposed to control doublon formation, both, spatially and temporally. 
Here, a novel mechanism that was recently proposed by us [Balzer \textit{et al.}, submitted for publication, arxiv:1801.05267] is verified by exact diagonalization and nonequilibrium Green functions (NEGF) simulations---fermionic doublon creation by the impact of energetic ions. We report the formation of a nonequilibrium steady state with homogeneous doublon distribution. 
%A physically intuitive picture is given in terms of an analytical model for a two-site system where the doublon formation is explained in terms of a two-fold passage of an avoided crossing (Landau-Zener picture).
The effect should be particularly important for strongly correlated finite systems, such as graphene nanoribbons,  
%in contact with a high-pressure plasma 
and directly observable with fermionic atoms in optical lattices.
We demonstrate that doublon formation and propagation in correlated lattice systems can be accurately simulated with NEGF. In addition to two-time results we present single-time results within the generalized Kadanoff-Baym ansatz (GKBA) with Hartree-Fock propagators (HF-GKBA). Finally we discuss systematic improvements of the GKBA that use correlated propagators (correlated GKBA) and a correlated initial state.
}

\maketitle   % please do not remove

\section{Introduction.}\label{s:intro}
The interaction of energetic charged particles with solid bodies is a phenomenon common to hot gases, plasmas, as well as astrophysical systems, including the solar wind and cosmic rays. When charged particles hit a solid surface, they deposit energy and momentum and may cause substantial surface modification the details of which strongly depend on the particle energy and the material properties. In low-temperature plasma physics, this process is routinely used to clean surfaces from adsorbates or modify them via sputtering, e.g. \cite{adamovich_2017_plasma} or as a source of secondary electrons \cite{bonitz_fcse_18}. On the other hand, ions impacting a solid can be used as a diagnostic tool of the electronic structure of the material by measuring the energy loss (or stopping power or stopping range) as a function of impact energy \cite{sigmund_springer_06}.

From the theory side, the interaction of ions with a solid surface has been studied with a variety of approaches including scattering theory \cite{nagy_pra_98} or uniform electron gas models \cite{pitarke_prb_95}. More recently, \textit{ab initio} simulations  of ion stopping based on time-dependent density functional theory (TDDFT) became available for metals \cite{quijada_pra_07}, semimetals \cite{ojanpera_prb_14} or boron nitride and graphene sheets \cite{zhao_2015_comparison} and other materials.
These simulations account primarily for valence electron excitation. Good results for the stopping
power of high energy ions in matter are also provided by the SRIM
code \cite{srim_10}
that uses the binary collision approximation in combination
with an averaging over a large range of experimental situations.
Thus presently two main questions remain open: I) how does the stopping power change in correlated materials and what is the effect of the correlation strength? II) How does the stopping power change when the system size is reduced or the geometry of the target is altered? And what is the role of electronic correlations in finite systems?

The motivation for these questions is fueled by the recent progress in nanostructured materials, clusters or finite nano-size systems. A particularly exciting example are finite honeycomb clusters or graphene nanoribbons (GNR). GNR hold the promise that they overcome the limitations of graphene arising from its semimetallic character. In contrast, GNR have been shown to have a finite bandgap $E_G(L)$ arising from the quantum confinement  \cite{Yang2007,Han2007}. Over a broad range of system widhts $L$, the band gap increases nearly proportional with $L^{-1}$ \cite{nakada_edge_1996}. Typical values for the bandgap are found to be $E_G \leq2.5\,\mathrm{eV}$ according to tight-binding and DFT calculations \cite{son_energy_2006}. Taking into account quasiparticle corrections results in a significantly larger gap of $E_G \leq5.5\,\mathrm{eV}$ \cite{yang_quasiparticle_2007}. In electronic structure measurements for GNRs on substrates bandgaps of $E_G \sim2.4-3.5\,\mathrm{eV}$ were found \cite{ruffieux_electronic_2012,sode_electronic_2015,wang_giant_2016}. 
The finite band gap makes the material semiconducting which is crucial for applications in electronics and optics.
Recent progress in synthetization methods of GNR \cite{Bai2009,Jiao2010,kimouche_ultra_narrow_2015}, has drastically increased the number of exciting experiments over the past few years \cite{jensen_ultrafast_2013,gierz_tracking_2015,senkovskiy_semiconductor_metal_2017,ivanov_role_2017}. Therefore, an accurate theoretical description of these systems in nonequilibrium and especially of their time-resolved correlation effects is needed. 
 
 However, finite graphene nanostructures, especially when driven out of equilibrium, are extremely complex, inhomogeneous systems that put high requirements on theory. %that attempts to describe them accurately. 
%A convenient theory has to describe finite systems including up to $100$ carbon atoms. Further, it has to take into account the finite overlap of the atomic orbitals and describe moderate electronic correlations. 
The two-dimensional geometry of the graphene honeycomb lattice has to be modeled, and the correlated nonequilibrium dynamics of the system have to be accurately described for up to several femtoseconds within a reasonable amount of computing time. 
Due to the limitations of time-dependent density functional theory to weakly correlated systems and the difficulties of density matrix renormalization group (DMRG) approaches to treat two-dimensional systems, nonequilibrium Green functions (NEGF) have emerged as the first choice to provide such a description. This method has recently been shown to accurately describe the dynamics of finite strongly correlated lattice systems, e.g. \cite{schluenzen_cpp16,schluenzen_prb16,schluenzen_prb17} where both two-time simulations and single-time dynamics within the generalized Kadanoff-Baym ansatz (GKBA \cite{lipavski_prb_86}) were presented \cite{hermanns_2014_hubbard}. Furthermore, in our recent work \cite{balzer_prb16,balzer_prl_18} we have shown that the NEGF approach is well capable to treat the correlated electron dynamics in lattice systems that is initiated by the impact of charged projectiles and, thus, is able to answer questions I. and II. that were raised above.

The goal of this article is to present recent results on NEGF simulations of finite correlated lattice systems with a particular focus on doublon creation and propagation following the impact of one or several charged particles.  We  also discuss how to include the description of charge transfer processes between projectile and target that is observed at low impact velocities.
Finally, we discuss theoretical issues that are related to the GKBA and to its extension to include correlated propagators.

The remainder of this paper is organized as follows. In Section \ref{s:model}
we introduce the Hubbard model and the description of the interaction of the charged projectile with the electronic system. This is followed, in Sec.~\ref{s:negf}, by a brief introduction into the NEGF approach and the GKBA and a discussion of its further improvements. The main results are presented in Sec.~\ref{s:results} and include numerical data from two-time and NEGF and GKBA simulations as well as analytical results for a representative two-site system, cf. Sec.~\ref{ss:dimer}. We conclude by presenting an embedding approach to treat the charge transfer between projectile and solid, in Sec.~\ref{s:embedding} and by an outlook, in Sec.~\ref{s:discussion}.

\section{Model.}\label{s:model}
We consider a 1D or 2D system with strong electronic correlations that is modeled by a Hubbard hamiltonian (\ref{eq:ham1}) with hopping amplitude $J$ [$\langle i,j\rangle$ denotes nearest neighbors] and onsite interaction strength $U$.
\begin{eqnarray}
\label{eq:ham1}
%H_\textup{p}&=-\frac{\hbar^2}{2m_\textup{p}}\nabla_\textup{p}^2+\frac{Z_\textup{p}e^2}{4\pi\epsilon_0}\sum_i\frac{Z_i\textup{e}^{-\kappa |\vec{r}_\textup{p}-\vec{R}_i|}}{|\vec{r}_\textup{p}-\vec{R}_i|}\,,\\
%\label{eq:ham2}
H_\textup{e}&=- J\sum_{\langle i,j\rangle,\sigma}
%T_{\langle ij\rangle}
c_{i\sigma}^\dagger c_{j\sigma} + U\sum_{i}\left(n_{i\uparrow}-\frac{1}{2}\right)\left(n_{i\downarrow}-\frac{1}{2}\right)&\nonumber\\
&
%\hspace{1pc}
-\frac{Z_\textup{p}e^2}{4\pi\epsilon_0}\sum_{i,\sigma}\frac{c_{i\sigma}^\dagger c_{i\sigma}}{|\vec{r}_\textup{p}(t)-\vec{R}_i|}
+ \sum_{\langle i,j\rangle,\sigma}W_{ij}(t)c_{i\sigma}^\dagger c_{j\sigma}
\,.
\end{eqnarray}
The strength of correlations is measured by the ratio $U/J$ and is typically in the range from $0$ to $10$.
The second line of Eq.~(\ref{eq:ham1}) contains the coupling of the lattice electrons located at coordinate ${\vec R}_i$ with a positively charged projectile of charge $Z_p$ that is treated classically (Ehrenfest dynamics) by solving Newton's equation for the trajectory $\textbf{r}_p(t)$ under the influence of all Coulomb forces with the lattice electrons. The final term allows to improve the model by accounting for modification of the hopping rates due to the projectile according to $W_{ij}(t)=\gamma [W_{ii}(t)+W_{jj}(t)]/2$, where $W_{ii}$ is the magnitude of the Coulomb potential  of the projectile at lattice site ``i'', and  $\gamma$ is a phenomenological parameter of the order unity \cite{balzer_prb16}. 

A good quality of the model (\ref{eq:ham1}) is achieved by using \textit{ab initio} input data for the model parameters by fitting them to DFT simulation results.
A further improved description of graphene-type finite size structures can be achieved via an \textit{extended Hubbard model} which is described in some detail in Ref.~\cite{joost_pss_18}.

\section{Nonequilibrium Green Functions Formalism.}\label{s:negf}
The method of nonequilibrium (real-time) Green functions is a very powerful approach to quantum many-body systems out of equilibrium, cf. Refs.~\cite{keldysh64,kadanoff-baym-book}.
The method successfully overcomes the limitations of the quantum Boltzmann equation, such as the restriction to times larger than the correlation time and fundamental problems such as failure for strongly correlated systems, incorrect conservation laws (e.g. conservation of kinetic energy instead of total energy) and relaxation toward an equilibrium state of an ideal gas (Fermi, Bose or Maxwell distribution) instead of the one of an interacting system, for a detailed discussion, see Refs.~\cite{bonitz_qkt,bonitz-etal.96pla,bonitz-etal.96jpb,bonitz96pla,kremp-etal.97ap}. An extensive overview on recent applications that span condensed matter physics, nuclear physics, laser plasmas etc. can be found in the proceedings of the PNGF conferences \cite{pngf1,pngf2,bonitz_pngf3,bonitz_jpcs_10,vanleeuwen_jpcs_13,verdozzi_jpcs16}.

\subsection{Basic concepts.}
The NEGF-method is formulated in second quantization (for textbook or review discussions, see e.g. Refs.~\cite{stefanucci_cambridge_2013,kadanoff-baym-book,schluenzen_cpp16}), in terms of creation (annihilation) operators $c^{\dagger}_{i\sigma}$ ($c_{i\sigma}$) for electrons in a single-particle orbital $|i\rangle$ with spin projection $\sigma$ that obey the standard fermionic anti-commutation relations. Below we will consider a spatially inhomogeneous lattice configuration where $i$ labels the spatial coordinates of individual lattice points.

The central quantity that determines all time-dependent observables is the one-particle NEGF 
%(we use $\hbar = 1$), 
\begin{equation}
G_{ij\sigma}(t,t') = 
%-\frac{i}{\hbar}
-\mathrm{i}\hbar
\langle T_{\cal  C}c_{i\sigma}(t)c^{\dagger}_{j\sigma}(t')\rangle\,,
    \label{eq:negf}
\end{equation}
where the expectation value is computed with the equilibrium density operator of the system, and times are running along the Keldysh contour $\cal C$, with $T_{\cal C}$ denoting ordering of operators on $\cal C$ \cite{keldysh64,bonitz_pss_18_keldysh}. %
The NEGF obeys
the two-time Keldysh-Kadanoff-Baym equations (KBE)~\cite{kadanoff-baym-book}
\begin{eqnarray}
\label{eq.kbe}
 &\sum_k &\left[\mathrm{i}\hbar\partial_t\delta_{ik}
 %+T_{\langle ik\rangle}
 -{\bar h}_{ik\sigma}(t)\right]G_{kj\sigma}(t,t')\\
 &&=\delta_{\cal C}(t-t')\delta_{ij}+\sum_{k}\int_{\cal C} ds\,\Sigma_{ik\sigma}(t,s)G_{kj\sigma}(s,t')\,,
\nonumber
\\
%\nonumber
 &\sum_k &G_{ik\sigma}(t,t')\left[-\mathrm{i}\hbar\overset{\leftarrow}\partial_{t'}\delta_{kj}
 %+T_{\langle ik\rangle}
 -{\bar h}_{kj\sigma}(t')\right]\\
 &&=\delta_{\cal C}(t-t')\delta_{ij}+\sum_{k}\int_{\cal C} ds\,G_{kj\sigma}(t,s)\Sigma_{ik\sigma}(s,t')\,,
\nonumber
\end{eqnarray}
where 
%the operators in the second equation act to the left and 
we do not consider spin changes. The hamiltonian ${\bar h}(t)$ contains kinetic, potential and mean field energy [including the projectile contributions in the second line of Eq.~(\ref{eq:ham1})], whereas correlation effects are contained in the selfenergy $\Sigma$.

For numerical applications the equations (\ref{eq.kbe}) for the Keldysh matrix Green function have to be rewritten for the correlation functions $G^\gtrless$:
\begin{align}
    \sum_l \left[ \mathrm{i}\hbar \partial_t \delta_{il} - {\bar h}_{il} (t) \right] G^{\gtrless}_{lj}(t,t') &= I^{(1)\gtrless}_{ij}(t,t') \label{KBE_1_coll} \, , \\
    \sum_l G^{\gtrless}_{il}(t,t') \left[- \mathrm{i}\hbar \overset{\leftarrow}\partial_{t'} \delta_{lj}  - {\bar h}_{lj} (t') \right] &= I^{(2)\gtrless}_{ij}(t,t') \label{KBE_2_coll} \, ,
\end{align}
with the collision integrals given by
\begin{align}
    &I^{(1)\gtrless}_{ij}(t,t') \coloneqq \label{eq:coll1}\\
    &\sum_l \int\limits_{t_\mathrm{s}}^\infty \mathrm{d} \overline{t}\, \left\{\Sigma^{\mathrm{R}}_{il}(t,\overline{t}) G^{\gtrless}_{lj}(\overline{t},t') + \Sigma^{\gtrless}_{il}(t,\overline{t}) G^{\mathrm{A}}_{lj}(\overline{t},t') \right\}\, ,\nonumber\\
    &I^{(2)\gtrless}_{ij}(t,t') \coloneqq \label{eq:coll2}\\
    &\sum_l \int\limits_{t_\mathrm{s}}^\infty \mathrm{d} \overline{t}\, \left\{G^{\mathrm{R}}_{il}(t,\overline{t}) \Sigma^{\gtrless}_{lj}(\overline{t},t') + G^{\gtrless}_{il}(t,\overline{t}) \Sigma^{\mathrm{A}}_{lj}(\overline{t},t')\right\}\,\nonumber,
\end{align}
where the retarded and advanced functions are given by 
\begin{align}
G^{R/A}_{ij}(t,t') &=\pm \Theta[\pm(t-t')]\left\{G^{>}_{ij}(t,t') - G^{<}_{ij}(t,t')\right\}\,,
\label{eq:gra-def}
\\
\Sigma^{R/A}_{ij}(t,t') &=\pm \Theta[\pm(t-t')]\left\{\Sigma^{>}_{ij}(t,t') - \Sigma^{<}_{ij}(t,t')\right\}.
\nonumber
\end{align}
Note that the correlation effects that are contained in the collision integrals $I^{1,2\gtrless}$ lead to memory effects, i.e. time integrations over the past, starting from a start time $t_s$. In most of the simulations presented below we will start at $t_s$ with an uncorrelated system and slowly switch on the interaction (``adiabatic switching'' \cite{hermanns_psc_12,hermanns_2014_hubbard}) which produces, at time $t_0$, a correlated ground state from which the excitation of the system starts. We return to the discussion of a correlated initial state in Sec.~\ref{ss:inicor}.

The system (\ref{eq.kbe})--(\ref{eq:gra-def})  is a closed set of equations for the dynamics of the NEGF once a selfenergy approximation $\Sigma[G]$ has been chosen. This issue is discussed in the following section.

\subsection{Selfenergies.}\label{ss:sigmas}
\begin{figure}[t]%
\centering
\includegraphics*[width=\linewidth]{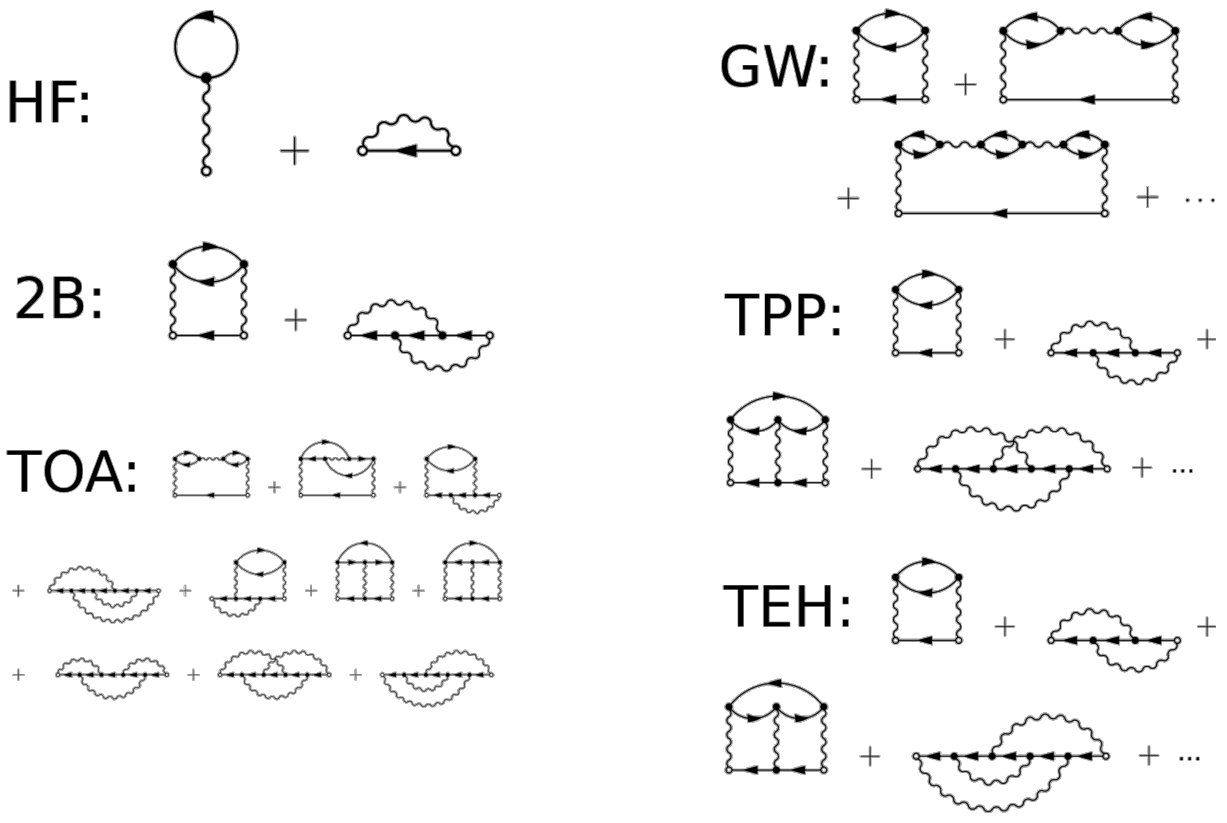}
\caption{%
  Feynman diagrams for the selfenergy approximations used in this work: Hartree-Fock (HF), Second order Born approximation (2B), particle-particle and particle-hole T-matrix (TPP, TEH), and third order approximation (TOA) which contains all diagrams with 3 interaction lines from GW, TPP and TEH.
  }
\label{fig:sigmas}
\end{figure}

In this work we use the following selfenergy approximations to account for the electron-electron interaction. We consider Hartree-Fock (HF) contributions (i.e. mean field, note that, for Hubbard systems, the Fock terms are absent) and correlation effects. The latter are described on the level of the second Born  (2B) and the T-matrix approximation (TM) where the former (latter) is adequate at weak (moderate) coupling \cite{schluenzen_prb16,schluenzen_prb17}. Moreover, we also consider the third-order approximation \cite{hermanns_phd_16} that includes all bubble and ladder-type diagrams to third order. The corresponding selfenergy diagrams are shown in Fig.~\ref{fig:sigmas}.

The KBE (\ref{eq.kbe}) are solved on the $t$-$t'-$plane as described in Refs.~\cite{balzer_2013_nonequilibrium,schluenzen_cpp16}. Due to the time integration involved in the collision integrals (memory) the numerical effort increases cubically with the simulation duration $T_{\rm tot}$. The effort is particularly high for the GW and T-matrix approximations since for the effective interaction, an additional integral equation has to be solved, e.g. \cite{schluenzen_cpp16}. One way to reduce the computational effort is the restriction to the propagation along the time diagonal via the generalized Kadanoff-Baym ansatz (GKBA), proposed in Ref.~\cite{lipavski_prb_86}. The GKBA reduces the computational effort of NEGF simulations with second order Born selfenergies from a scaling $\sim T_{\mathrm{tot}}^3$ with the total simulation duration to $\sim T_{\mathrm{tot}}^2$ as was confirmed in Ref.~\cite{hermanns_psc_12}.
The GKBA has the additional attractive feature that it reduces the degree of selfconsistency in the NEGF simulations \cite{hermanns_2014_hubbard} and ``cures'' the artificial damping behavior of two-time simulations observed in small systems at very strong excitation \cite{von_friesen_2010_kadanoff}, for computational aspects, see also Ref.~\cite{schluenzen_prb17_comment}.

\subsection{Generalized Kadanoff-Baym ansatz. Extension to correlated propagators.}\label{ss:gkba}
Instead of propagating the Green functions in the two-time plane one can perform a propagation along the diagonal, $T=(t+t')/2$, only. The equation for $G^<$ is a commutator equation -- the first equation of the BBGKY-hierarchy for the reduced density operators \cite{bonitz_qkt}:
\begin{eqnarray}
 \i \hbar \partial_TG_{ij}^<(T,T) &=& [{\bar h}(T), G^<(T,T)]_{ij} + I_{ij}(T),
 \label{eq:kbe-diagonal}
 \\[1ex]
 [ A , B ]_{ij} &=& \sum_{k} \left(A_{ik}B_{kj}-B_{ik}A_{kj}\right)\,,
 \nonumber\\
I_{ij}(T)&=&  \sum_{k} \int_{t_0}^T d\overline{t} \bigg\{ \Sigma_{ik}^>(T,\overline{t}) G_{kj}^<(\overline{t},T) 
\nonumber\\
&& \qquad -
\Sigma_{ik}^<(T,\overline{t}) G_{kj}^>(\overline{t},T) 
%\right.
%\nonumber\\
%  & & \qquad - \qquad                                %\left.
%  G_{ik}^>(T,\overline{t}) \Sigma_{kj}^<(\overline{t},T)                                                                 \nonumber\\                                                  &&\qquad + \qquad G_{ik}^<(T,\overline{t}) \Sigma_{kj}^>(\overline{t},T) 
+ \mathrm{h.c.}\bigg\}
    \,.
\label{eq:idiagonal-def}
 \end{eqnarray}
%where ${\bar h}(t)$ is the single-particle hamiltonian including all external potentials and the Hartree-Fock energy.
To compute the collision integral $I$, the Green functions $G^\gtrless(t,t')$ are required also away from the diagonal. In fact, due to the symmetry $G^\gtrless_{ij}(t,t') = - [G^\gtrless_{ji}(t',t)]^*$ values for $t\ge t'$ are sufficient. With the GKBA the following ``reconstruction'' approximation is made \cite{lipavski_prb_86}
\begin{eqnarray}
 G_{ij}^\gtrless(t,t') =  \i \hbar \sum_k 
% \left( 
 G_{ik}^\mathrm{R}(t,t') G_{kj}^\gtrless(t',t'), \quad t\ge t'\, 
% - G_{ik}^\gtrless(t,t) G_{kj}^\mathrm{A}(t,t') 
% \right)
 \,,\label{GKBA}
\end{eqnarray}
and with $G^\gtrless(t,t')$ also $\Sigma^\gtrless(t,t')$ are known. While the diagonal value $G_{kj}^\gtrless(t',t')$ is available from the solution of Eq.~(\ref{eq:kbe-diagonal}), the retarded function has to be provided as an external input. Among the different approaches in macroscopic systems we mention the use of ideal propagators (``Free GKBA'' or FGKBA), quasiparticle propagators that are exponentially decaying as a function of $|t-t'|$ (QP-GKBA) which have been used extensively in semiconductor optics and transport, in particular, by the groups of Haug, Banyai and Jahnke, e.g. \cite{banyai_prl_95,haug_2008_quantum,lorke_jpcs_06,seebeck_prb_05} and references therein. For strong field physics in semiconductors and laser plasmas the gauge-invariant FGKBA has been introduced \cite{haug_2008_quantum,kremp_99_pre,bonitz_99_cpp,haberland_01_pre}. The GKBA has also been used with propagators taken from a full two-time simulation (2t-GKBA) in Ref.  \cite{kwong-etal.98pss} which confirmed the good quality of the ansatz (\ref{GKBA}).
A revival of the interest in the GKBA occured with the NEGF study of finite systems about a decade ago, e.g. \cite{stefanucci_cambridge_2013} and references therein. Here very good results were obtained with Hartree-Fock propagators (HF-GKBA) \cite{balzer_2013_nonequilibrium,hermanns_jpcs13,balzer_jpcs13,bonitz_cpp13}. 

\begin{figure}[h]%
\includegraphics*[width=\linewidth]{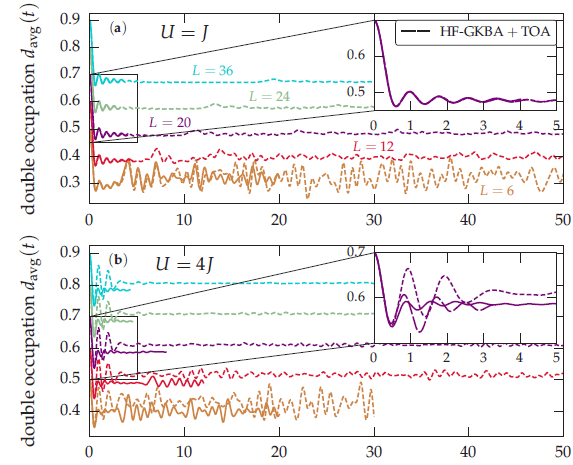}
\caption{%
Benchmarks of the HF-GKBA against DMRG for a 1D charge density wave state of doublons (site occupations alternate $2,0,2\dots$).
System-size dependence and long-time evolution of the average double occupancy, Eq.~(\ref{eq:d-av}), for (a) $U=J$ and (b)
$U=4J$ and chain lengths $L=6, 12, 20, 24, 36$.  Full  lines:  DMRG, short dashes: HF-GKBA+TMA. The insets, in addition, show HF-
GKBA+TOA results (long dashes). For better visibility, curves for
different $L$ are shifted vertically by 0.1.
 After Ref.~\cite{schluenzen_prb17}.}
\label{fig:negf-dmrg}
\end{figure}

While earlier studies used the GKBA together with lowest order correlated selfenergies (second Born approximation) we recently demonstrated that the HF-GKBA can also be successfully used together with mored advanced approximations such as the T-matrix, GW and third-order selfenergies, cf. Sec.~\ref{ss:sigmas}.
The most thorough test of the HF-GKBA (and of two-time NEGF simulations), so far, was performed in Ref.~\cite{schluenzen_prb17} by benchmarks against quasi-exact DMRG simulations for 1D systems which are summarized in Fig.~\ref{fig:negf-dmrg}. For weak and moderate coupling very good agreement with DMRG was obtained, if the HF-GKBA was combined with the adequate selfenergy: second order Born for $U/J\le 1$ and T-matrix for $U/J\le 4$ at weak (or high) filling. Around half filling the third order approximation showed the best behavior. This agreement is observed for all observables including densities and energies and even for very sensitive quantities such as the average double occupation, Eq.~(\ref{eq:d-av}), that is shown in Fig.~\ref{fig:negf-dmrg}. While the NEGF simulations are more efficient the DMRG at weak and moderate coupling (cf. the accessible simulation durations in Fig.~\ref{fig:negf-dmrg}), 
for strong coupling, $U=10$, in contrast to DMRG, no NEGF simulations were possible, indicating complementary applicability ranges of the two methods \cite{schluenzen_prb17}. In addition, NEGF have the remarkable advantage of being completely flexible in terms of system dimensionality and geometry which makes them an ideal approach to treat finite correlated systems such as GNR.

Despite the success of the HF-GKBA, it also has problems. While it removes most of the over-damping artifacts of two-time NEGF simulations (see above), it often underestimates the damping present in the exact dynamics and does not correctly reproduce the high-frequency features, cf. Fig.~\ref{fig:negf-dmrg}. Also, due to the HF-propagators, the spectral function produced by the HF-GKBA is uncorrelated. There have been early attempts to modify the free propators by an exponential damping, $G^R \sim e^{-\gamma |t-t'|}$ (cf. QP-GKBA above). However this choice of propagators violates 
energy conservation [as opposed to the FKGBA and HF-GKBA] due to a very slow ($1/\omega^2$) decay of the propagators in frequency space. This behavior was improved in Ref.~\cite{bonitz-etal.99epjb} by the use of non-Lorentzian damping factors, $G^R \sim 1/\cosh^\alpha{[\omega (t-t')]}$, where $\omega$ is a characteristic frequency (phonon or plasmon frequency) and $\alpha$ is a positive fit parameter,  but energy conservation is still violated.
For a recent discussion of the reconstruction problem, see Ref.~\cite{spicka_ijmpb_14}.

Here we outline a systematic approach towards an improved version of the GKBA that goes beyond the HF-GKBA. The idea is to start from the equation of motion for the retarded propagators (Dyson equation)
\begin{align}
 G_{ij}^\mathrm{R}(t,t') &= G_{\mathrm{HF},ij}^\mathrm{R}(t,t') + \sum_k\int\limits_{t'}^{t}d\tilde{t}\,G_{\mathrm{HF},ik}^\mathrm{R}(t,\tilde{t})\, \tilde{I}^R_{kj}(\tilde{t},t')\,.
\nonumber
\\
     \tilde{I}^R_{ij}({t},{t'})&= \sum_k\int_{{t'}}^{{t}} dt''\,\tilde{\Sigma}^R_{ik}({t},t'')\,G^R_{kj}(t'',{t}')\,,
    \label{eq:col-int-gr-gkba}
% \label{eq:grsol-gkba}
 \end{align}
where $\tilde{\Sigma}$ is a conserving selfenergy that may be different from the one used in $I_{ij}$ \cite{hermanns_2014_hubbard}. 
Since our main goal is to improve the single-time simulations beyond the HF-GKBA and to include damping effects, we may regard correlation effects in the GKBA as a small corrections to $G_{\mathrm{HF},ij}^\mathrm{R}(t,t')$. While the HF-GKBA corresponds to the neglect of the integral in (\ref{eq:col-int-gr-gkba}), an approximate treatment of the integral will be called \textit{correlated GKBA} (C-GKBA). For this we propose several  approximations that are listed in increasing order of accuracy, assuming that $\tilde{\Sigma}$ corresponds to weak correlations, i.e. small $\tilde{U}/J$:
\begin{enumerate}
\item[(a) ] replacement of all propagators in the integral (\ref{eq:col-int-gr-gkba}) by ideal propagators, $G^R \to G^R_{\mathrm{id}}$;
\item[(b) ] replacement of all propagators in the integral (\ref{eq:col-int-gr-gkba}) by HF propagators, $G^R \to G^R_{\mathrm{HF}}$. The result $G^{R,(1)}$ can be understood as first step of an iteration series that starts with $G^{R,(0)}\equiv G^R_{\mathrm{HF}}$;
\item[(c) ] higher order iterations, $G^{R,(l)}$, $l\ge 2$, that use $G^{R,(l-1)}$ in the integral term;
\item[(d) ] linearization of the collision integral in the correlated $G^R$. This means, products of retarded functions are replaced according to $G^R_{ik}G^R_{kj} \to G^R_{\mathrm{HF},ik}G^R_{kj}+G^R_{ik}G^R_{\mathrm{HF},kj}$ and similarly, for more complex products;
\item[(e) ]  2t-GKBA: exact solution of the Dyson equation for $G^R(t,t')$ \cite{kwong-etal.98pss}, see above;
\end{enumerate}
%A first improvement over the  is then achieved by making, everywhere in the integral, the replacement $G_{ij}^\mathrm{R}(t,t') \to G_{\mathrm{HF},ij}^\mathrm{R}(t,t')$. Alternatively, assuming that correlations are weak, we may also use $G_{ij}^\mathrm{R}(t,t') \to G_{\mathrm{id},ij}^\mathrm{R}(t,t')$, neglecting the mean field (which is of order $U/J$). This can be considered as the first iteration to the full solution. Further improvements can be derived by higher order iterations. Note that the full solution of the Dyson equation (\ref{eq:col-int-gr-gkba}) will not only involve the functions $G^{R/A}$ but also $G^\gtrless$, away from the time diagonal, which can again be reconstructed via the GKBA. 
Note that the Dyson equation (\ref{eq:col-int-gr-gkba}) for $G^R$ is not closed since the selfenergy $\tilde{\Sigma}^R$, in general, also contains $G^\gtrless$. However, in the spirit of perturbation theory we can always reconstruct $G^\gtrless$ via $G^{R/A}$ applying again the GKBA (\ref{GKBA}).

This is a systematic scheme to incorporate correlations in the propagators. The drawback of the C-GKBA is, of course, that the evaluation of the integral term in Eq.~(\ref{eq:col-int-gr-gkba}) is costly, scaling as $T_{\rm tot}^3$. However, this effort is warranted by the expected improved accuracy of the observables and spectral properties as compared to two-time NEGF simulations, on the one hand, and HF-GKBA results, on the other. The analytical and numerical properties of the C-GKBA are presently under investigation. Finally, we note that recently also improvements that take into account corrections beyond the GKBA have been studied for stationary transport problems by Kalvova \textit{et al.} \cite{kalvova_epl_18}. A modified reconstruction problem where the GKBA is applied also to the off-diagonal propagation (``extended GKBA'') was recently proposed by Hopian \textit{et al.} \cite{hopian_2018_phd,verdozzi_gkba-note} but the relation to the original reconstruction scheme of Ref.~\cite{lipavski_prb_86} remains open.\\

\subsection{Initial correlations for NEGF and GKBA. Restart capability.}\label{ss:inicor}
Until now we have only considered situations where, at the ``initial'' time where the evolution starts, the system is uncorrelated. This is, of course, a special case. In general, at this time, the system may be characterized by non-vanishing pair correlations $c_{12}$ which may have a profound effect on the dynamics. The generalization of the KBE to include finite initial correlations goes back to Danielewicz \cite{DANIELEWICZ_84_ap2} who derived a collision integral $I^{\mathrm{IC}}$ that is due to $c_{12}$. Alternative derivations have been given by Kremp \textit{et al.} who also derived initial correlation contributions to the selfenergy \cite{kremp_99_pre,semkat_00_jmp}. In these papers also numerical results were given that demonstrate the effect of initial correlations. Text book discussions can be found in Refs.~\cite{kremp-springer,bonitz_qkt,stefanucci_cambridge_2013}. Despite these early results and similar theoretical and numerical results for density operators, e.g.~\cite{bonitz_qkt}, numerical results for the GKBA have not been proposed so far. Only recently, two papers appeared that presented solutions for this problem \cite{verdozzi_gkba-note,karlsson_gkba18}.

Here we present an alternative approach that is based on Ref.~\cite{semkat_cpp_03} that provides a complementary and more general view on this issue.  In Eq.~(\ref{eq.kbe}) we introduced, on the right-hand side, the collision integral that involves the correlation selfenergy or, alternatively, the correlation part of the two-particle Green function $G^{(2)}$
\begin{align}
&    \int d2 V(1-2)\,G^{(2)}(12,1'2^+) = \int_{C}d{\bar 1}\,\Sigma(1,{\bar 1})\,G({\bar 1},1')
    \label{eq:ms-decoupling}
    \\
&\qquad  =  I(1,1';-\infty) \equiv I(1,1';t_0) + I^{\mathrm{IC}}(1,1')\,.
    \label{eq:iic}
\end{align}
Here the third argument of $I$ explicitly denotes the initial moment of the time evolution. When the evolution starts at $-\infty$, the system is assumed to be uncorrelated initially and, due to collisions, correlations are being build up until, at a finite time $t_0$ they reach a value $c(t_0)$. This can be real dynamics driven by an external excitation. Alternatively, if one is interested in a correlated initial state, the evolution from $-\infty$ to $t_0$ can be generated ``artificially'' by adiabatically switching on the interaction, starting from an uncorrelated state, e.g. \cite{hermanns_2014_hubbard} or via including an imaginary track into the Keldysh contour, e.g. \cite{balzer_2013_nonequilibrium,stefanucci_cambridge_2013}. Even though the start of the dynamics is, in practice, set to a finite value, $-\infty \to t_-$ with $c(t_-)=0$, both scenarios involve a time integration over the past in the r.h.s. of Eq.~(\ref{eq:ms-decoupling}) which is computationally costly, in particular for long propagation times.

This expensive time integration from $t_-$ to $t_0$ can, in fact be avoided in many cases \cite{kremp_99_pre,karlsson_gkba18} as we show now. 
The r.h.s. of Eq.~(\ref{eq:iic}) indicates that the collision integral can be identically rewritten as a scattering integral $I$, in which the evolution starts at $t_0$, plus an additional collision integral $I^{\mathrm{IC}}$ that contains the initial correlations $c(t_0)$, for a detailed discussion, see Ref. \cite{semkat_cpp_03}. In that reference explicit results for a homogeneous system were given. Using the momentum representation (plane wave basis) the additional collision integral becomes
\begin{align}
& I^{\mathrm{IC}}_{p_1}(t,t')=  -2\mathrm{i}\hbar^5 {V_0}
\sum\limits_{p_2 {\bar p}_1{\bar p}_2} V_{p_1-\bar{p}_1}\delta_{p_1+p_2,{\bar p}_1+{\bar p}_2} \times
\label{eq:iictt-born}\\
& G^R_{{\bar p}_1}(t,t_0)G^R_{{\bar p}_2}(t,t_0)c_{{\bar p}_1,{\bar p}_2,p_1,p_2}(t_0)
G^A_{{\bar p}_1}(t_0,t')G^A_{{\bar p}_2}(t_0,t')\,,
\nonumber    
\end{align}
where $V_0$ is the volume. This is the first crucial step and one realizes that Eq.~(\ref{eq:iictt-born}) does, indeed, not contain a time integral. The second important step is to derive the initial correlation function $c(t_0)$. This is done by going back to the connection between the two-particle Green function and the selfenergy, Eq.~(\ref{eq:ms-decoupling}), and to specialize this to the desired time moment, $t=t' \to t_0$. This leads to the following relation 
\begin{align}
& I^{\mathrm{IC}}_{p_1}(t_0,t_0)=  -2\mathrm{i}\hbar {V_0}
\sum\limits_{p_2 {\bar p}_1{\bar p}_2} V_{p_1-\bar{p}_1}\delta_{p_1+p_2,{\bar p}_1+{\bar p}_2} \times
\nonumber    
\\
&\qquad\qquad\qquad\qquad\qquad\qquad c_{{\bar p}_1,{\bar p}_2,p_1,p_2}(t_0) 
\label{eq:iictt-consistency}
\\
%\label{eq:iictt-consistency}
&=\int_{t_-}^{t_0} d{\bar t} \big\{ \Sigma^>_{p_1}(t_0,{\bar t})G^<_{p_1}({\bar t},t_0)
-\Sigma^<_{p_1}(t_0,{\bar t})G^>_{p_1}({\bar t},t_0)
\big\}\,
\nonumber
\end{align}
which constitutes an equation for the matrix $c(t_0)$ in terms of the selfenergy and the correlation functions built up from the uncorrelated state at $t_-$.
An explicit result for $c(t_0)$ can be obtained for direct second order Born selfenergies (first 2B diagram in Fig.~\ref{fig:sigmas}), for ${\bar p}_1+{\bar p}_2=p_1+p_2$ (the other matrix elements are equal to zero),
\begin{align}
& c_{{\bar p}_1,{\bar p}_2,p_1,p_2}(t_0) =\frac{\mathrm{i\hbar}}{V_0}\int_{t_-}^{t_0}d{\bar t}\,V_{p_1-\bar{p}_1}\times
    \label{eq:result_c_born}
\\
& \big\{
G^>_{{\bar p}_1}(t_0,{\bar t})G^>_{{\bar p}_2}(t_0,{\bar t})G^<_{p_1}({\bar t},t_0)G^<_{p_2}({\bar t},t_0) - (> \leftrightarrow <)
\big\}\,,
\nonumber
\end{align}
which was presented in Ref.~\cite{semkat_cpp_03} for the general case of NEGF propagation in the two-time plane. 

Expression (\ref{eq:result_c_born}) is immediately rewritten for the case of propagation along the time diagonal within the GKBA scheme, cf. Sec.~\ref{ss:gkba}, by replacing the functions $G^\gtrless$ via (\ref{GKBA}),
\begin{align}
& c^{\mathrm{GKBA}}_{{\bar p}_1,{\bar p}_2,p_1,p_2}(t_0) =\frac{\mathrm{i\hbar}}{V_0}\int_{t_-}^{t_0}d{\bar t}\,V_{p_1-\bar{p}_1}\times
    \label{eq:result_c_born-gkba}
\\
& \qquad
G^R_{{\bar p}_1}(t_0,{\bar t})G^R_{{\bar p}_2}(t_0,{\bar t})G^A_{p_1}({\bar t},t_0)G^A_{p_2}({\bar t},t_0) 
\nonumber\\[1ex]
&\times
\big\{f^>_{{\bar p}_1}(t_0)f^>_{{\bar p}_2}(t_0)f^<_{p_1}(t_0)f^<_{p_2}(t_0)
- (> \leftrightarrow <)
\big\}\,,
\nonumber
\end{align}
where $f^<(t_0)$ is the Wigner function of the initial state, and $f^>=1\pm f^<$. If HF propagators are chosen this agrees with the result of Ref.~\cite{karlsson_gkba18}, but improved propagators can also be used, as was discussed in Sec.~\ref{ss:gkba}. 
Another approach is to derive $c(t)$, Eq.~(\ref{eq:result_c_born}), from the Bethe-Salpeter equation for $G^{(2)}$. For any choice of the selfenergy $\Sigma$ it is possible to find the functional $G^{(2)}[G]$, as was explicitly demonstrated for the Born approximation in Ref.~\cite{bonitz_jpcs_13}. With the GKBA this also provides the result for $c^{\mathrm{GKBA}}(t_0)$, Eq.~(\ref{eq:result_c_born-gkba}).
In fact, the result for $c^{\mathrm{GKBA}}(t_0)$ with HF propagators does not require NEGF input at all. It follows directly from density operator theory within the single-time BBGKY-hierarchy where it has been computed for a variety of many-particle approximations including second order Born, T-matrix \cite{kremp-etal.97ap,semkat_00_jmp} or GW approximation \cite{bonitz_qkt}.  

Finally we note that this approach of computing the quantum dynamics within the two-time NEGF or single-time GKBA scheme by starting from a correlated state at a finite time $t_0$ has another important application. Indeed, the pair correlation $c(t_0)$ is not necessarily that of the ground state or the equilibrium state,  but it is arbitrary, as long as it fulfills condition (\ref{eq:iictt-consistency}) as was shown in Ref.~\cite{semkat_cpp_03}. For example, it can be the correlations that have been built up during a previous real dynamics, for $t\le t_0$, and which can be used to restart (continue) the evolution, for $t\ge t_0$, cf. Ref.~\cite{semkat_cpp_03}. This is possible in cases when a unique solution of Eq.~(\ref{eq:iictt-consistency}) for the entire matrix of $c$ exists.\\

\subsection{NEGF-Ehrenfest approach to ion stopping.}
Let us now come back to the problem of 
 ion stopping and the associated electronic correlation effects in finite graphene-type clusters that we discussed above in Secs.~\ref{s:intro} and \ref{s:model}. For the numerical analysis,
 we use the Kadanoff-Baym equations (\ref{eq.kbe}) with the electronic hamiltonian (\ref{eq:ham1}). The impacting ion acts as a time-dependent external attractive potential for all electrons. This potential is sharply peaked as a function of time, reaching its maximum (negative) value when the projectile traverses the graphene layer.
The energy loss of the ion is treated classically via solution of Newton's equation (Ehrenfest dynamics).
Processes of charge transfer between target and projectile which are important at low impact velocities will be considered separately, in Sec.~\ref{s:embedding}.

From the NEGF all time-dependent single-particle observables can be computed according to 
\begin{equation}
    \langle \hat{A} \rangle (t) = -\mathrm{i}\hbar \sum_{ij} A_{ij}G_{ji}^<(t,t)\,,
\end{equation}
including the single-particle energy and the site-resolved density, $n_{i\sigma}=\langle \hat{n}_{i\sigma}(t)\rangle$. 
Another important quantity is the time-resolved photoemission spectrum \cite{eckstein-pes_2008}
\begin{align}
  A^<(\omega, T) = -\mathrm{i} \hbar \sum_{i} &\int \mathrm{d}t\, \mathrm{d}t' \, \mathcal{S}_\kappa(t-T) \mathcal{S}_\kappa(t'-T)\nonumber\\
  &\times \mathrm{e}^{\mathrm{-i} \omega (t-t')}  G_{ii}^<(t,t')\,, \label{eq:photoemission}
\end{align}
which measures the occupied states of the system. It allows for a direct comparison with time-resolved (pump-probe) photoemission  experiments where $\mathcal{S}$ mimicks a Gaussian  probe pulse of width $\kappa$,
\begin{equation}
  \mathcal{S}_\kappa(t) = \frac{1}{\kappa\sqrt{2\pi}} \exp\left(-\frac{t^2}{2\kappa^2}\right)\, .
%  \label{probe_pulse}
\nonumber
\end{equation}

The energy exchange between projectile and the cluster can be computed from the increase of the total energy of the electrons or, equivalently, from the energy loss of the projectile, 
\begin{equation}
    S_e = m_p \frac{{\dot r}^2_p(t\to -\infty)}{2} - m_p \frac{{\dot r}^2_p(t\to +\infty)}{2},
\label{eq:se}
\end{equation}
which is just the difference of kinetic energies far away from the target before and after the impact. With this we assume that the interaction between different projectiles or with a surrounding plasma medium is negligible. Further, we do not resolve internal degrees of freedom of the projectile.
Also two-particle expectation values such as the correlation energy and the double occupation $d_i$ are accessible in the NEGF approach taking advantage of the two-time information in $G$ and $\Sigma$. Thus we compute the expectation value of the site-resolve doublon number, its cluster-average 
and the long-time limit of the latter, after passing of the projectile, according to
\begin{align}
\quad d_i(t) &=
\langle \hat{n}_{i\uparrow}(t)\hat{n}_{i\downarrow}(t)\rangle
\nonumber\\
&=
-\frac{\mathrm{i\hbar}}{U}\sum_k \int_{\cal C} ds\, \Sigma_{ik}(t,s)G_{ki}(s,t)\,,
\label{eq:mean-di}
%\label{eq:n-d}
\\
    d_\textup{av}(t) &= \frac{1}{L} \sum\limits_{i=1}^{L}d_i(t), 
\;\;     d^\infty_\textup{av} =\lim_{t\to \infty}\tfrac{1}{\Delta t}\int\limits
_t^{t+\Delta t}\!d{\bar t} \, d_\textup{av}({\bar t})\,.
%\langle  %\chat{n}_{i\uparrow}\chat{n}_{i\downarrow}\rangle
\label{eq:d-av}
\end{align}

\section{Results.}\label{s:results}
We now turn to the results for the time-resolved coupled electron-projectile dynamics. A detailed investigation has been presented in Ref.~\cite{balzer_2016_stopping,balzer_prl_18} some results of which are briefly summarized here and complemented with additional data.
For small clusters, $L\le 12$, we have performed exact diagonalization calculations whereas for larger systems we solved the Keldysh-Kadanoff-Baym equations (\ref{eq.kbe}) for the NEGF. In the latter case the accuracy of the results is determined by the choice of the selfenergy $\Sigma$. In this paper we present simulations within the second order Born approximation using the HF-GKBA, cf. Sec.~\ref{ss:gkba} and selected data with more advanced selfenergies that were introduced in Sec.~\ref{ss:sigmas}.
Prior to the NEGF simulations we have performed detailed numerical convergence tests that include particle number and energy conservation \cite{schluenzen_prb17_comment} and time reversibility \cite{scharnke_jmp17,bonitz_cpp18}. In addition, for small systems we have performed tests against exact diagonalization calculations.
Further tests of the present code (T-matrix selfenergy) include comparisons with cold atom experiments \cite{schluenzen_prb16} where excellent agreement was found. Finally we mention
extensive benchmarks against density matrix renormalization group (DMRG) calculations \cite{schluenzen_prb17}, a typical example -- for the GKBA -- was shown above in Fig.~\ref{fig:negf-dmrg}. An important outcome of the benchmarks of Ref.~\cite{schluenzen_prb17} was that the exact result is often enclosed between the two-time simulations and the HF-GKBA. 
From this we can conclude that the present NEGF stopping simulations are reliable and have predictive power.

\begin{figure}[h]%
\includegraphics*[width=\linewidth]{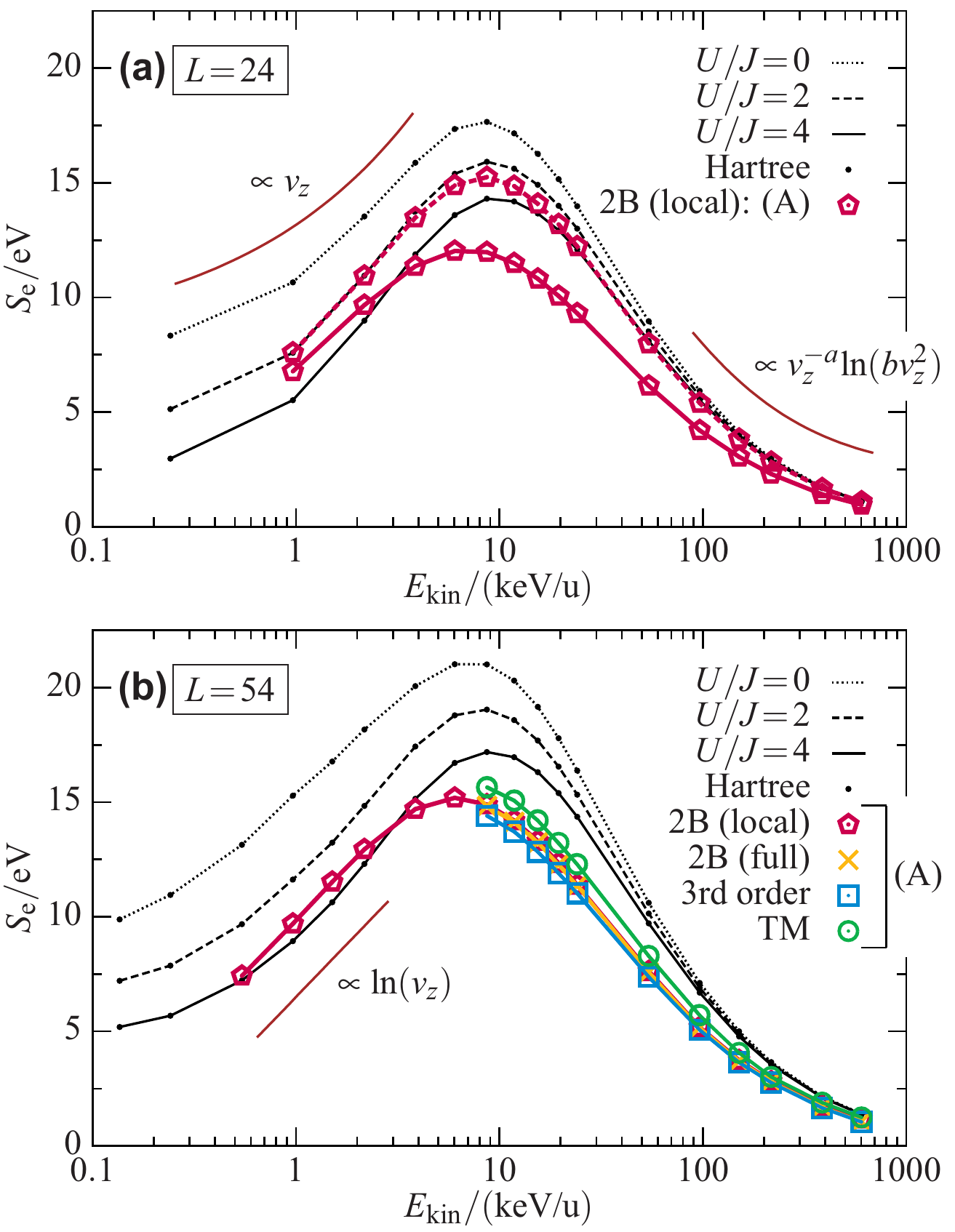}
\caption{%
Energy loss of a proton with initial energy $E_{\rm kin}$ during passage through a honeycomb cluster [cf. Fig.~\ref{fig:cluster}] of size $L=24$ (top) and $L=54$ (bottom). From Ref.~\cite{balzer_prb16}.}
\label{fig:stopping-curve}
\end{figure}
\subsection{Energy loss of the projectile.}\label{ss:se}
Let us start with the total energy loss of the projectile, Eq.~(\ref{eq:se}), as a function of impact energy which is shown in Fig.~\ref{fig:stopping-curve}, for the case of a proton. The overall behavior is well-known: the energy loss vanishes, both, for very low and very high impact energies. An optimum projectile-target interaction is observed at intermediate impact energies, in the range of several keV per mass unit $u$. The decrease at large energies is due to the reduced interaction duration and is consistent with the standard non-relativistic Bethe formula, e.g.~\cite{sigmund_springer_06}. Not surprisingly, here correlations in the material have very little influence which can be seen in the convergence of the curves for different $U$. In the opposite limit, the energy available for transfer to the target is small. At the same time, in the range left of the maximum the influence of the target properties on the energy loss is significant: here the curves for different coupling strength $U$ differ significantly. 

This overall trend of the energy loss (stopping power) is well reproduced with our NEGF simulations, and the results agree well with other approaches, such as TDDFT and the SRIM code, at high energies. On the other hand, in the low energy range the situation is less clear. One reason is that, previously, most attention focused on high-energy particle beams or hot plasmas. Only more recently low projectile energies in the range of several hundred or tens of eV attracted interest because this is the typical energy range in low-temperature plasmas and surface physics, e.g. \cite{bonitz_fcse_18}. In this range, correlation effects in the target (the value of $U/J$ in our model) play a crucial role, and also size and geometry effects are expected to be relevant. The influence of system size is clearly seen in our simulations, compare parts (a) and (b) of Fig.~\ref{fig:stopping-curve}: with increasing size of the cluster more electrons are excited by the projectile and, hence, the energy deposition, $S_e$, grows.

With the increasing role of correlations, also the requirements for theory increase. For NEGF simulations, this means that the proper choice of the selfenergy becomes important, whereas, at high impact energy, the difference between different selfenergy approximations is rather small, cf. Fig.~\ref{fig:stopping-curve} (b). At the same time, reducing the impact energy increases the interaction time and, thus, also the simulation duration in our nonequilibrium approach grows rapidly. For this reason, in the range of $1 keV/u$ and below, so far, mostly local second order Born simulations (assuming $\Sigma_{ij}\sim \Sigma_i\delta_{ij}$) were performed. A comparison to mean field (Hartree) simulations clearly signals the importance of correlations for the stopping for strongly correlated materials, cf. curves for $U/J=4$ in Fig.~\ref{fig:stopping-curve}(a).

\subsection{Ion impact induced doublon excitation.}\label{ss:doublons}
A particularly interesting observation is that the deviation of the correlated simulations from the mean field result changes sign. While for high energy, correlations seem to lower the energy deposition, at impact energies below approximately $3keV/u$, correlation effects enhance the stopping power. This is a surprising effect, and one may speculate that this is due to an increase of the correlation energy. To verify this hypothesis we analyze, in the following, the doublon number, Eq.~(\ref{eq:mean-di}), that is induced by the projectile. In fact, the total number of doublons or its cluster average, $d_{\mathrm{av}}$, Eq.~(\ref{eq:d-av}), minus the mean field result, 
\begin{eqnarray}
d^{\mathrm{H}}_i= \langle \hat{n}_{i\uparrow}(t)\rangle \langle\hat{n}_{i\downarrow}(t)\rangle = {n}_{i\uparrow}(t){n}_{i\downarrow}(t)\,,
\label{eq:di-hartree}
\end{eqnarray}
%
%\begin{figure}[h]%
%\includegraphics*[width=\linewidth]{figures/...}
%\caption{%
%\textbf{Spectral function before, during and after the projectile for low impact velocity and different values of $U$ and different system sizes $L$.
%}
%}
%\label{fig:stopping-spectral-fct}
%\end{figure}
%
\begin{figure}[htb]%
  \sidecaption
  \includegraphics*[width=.2\textwidth,height=3.5cm]{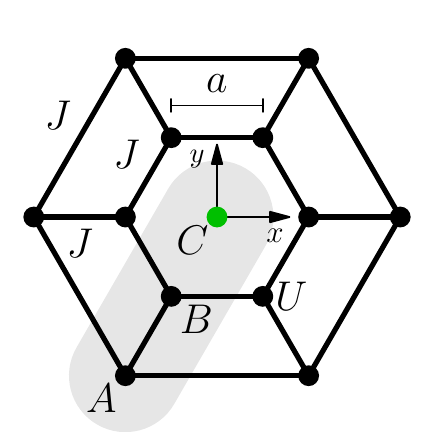}%
  \caption[]{%
    Sketch of a honeycomb cluster of $L=12$ sites and distance between sites $a$, showing the hopping and on-site interaction parameters in the hamiltonian (\ref{eq:ham1}). The dimer model of Sec.~\ref{ss:dimer}, consists of the representative sites A and B.}
    \label{fig:cluster}
\end{figure}
  \begin{figure}[h]
  \begin{center} 
%    \hspace{-0.cm}\includegraphics[width=0.2\textwidth]{figures/lattice.pdf}
  \hspace{-0.cm}\includegraphics[width=0.485\textwidth]{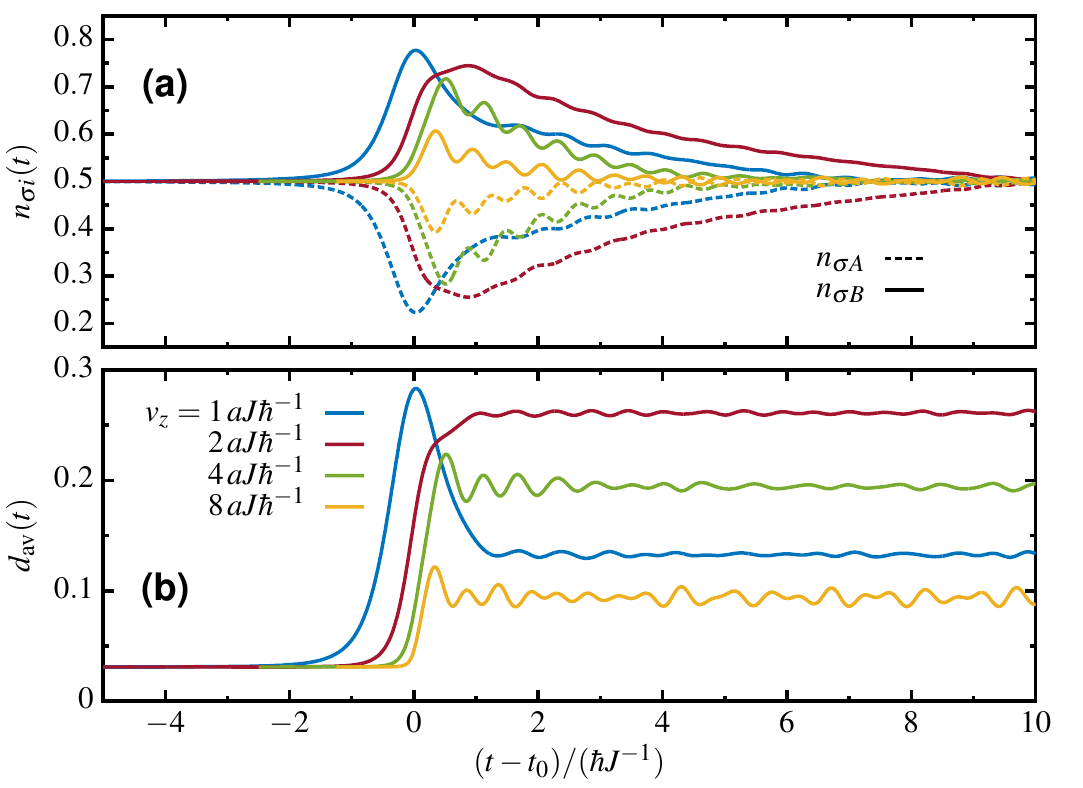}
  \end{center}
  \vspace{-.40cm}
  \caption{Time-dependent response of a strongly correlated finite honeycomb cluster of Fig.~\ref{fig:cluster} for $U/J=10$ to a charged projectile with charge $Z=2$ penetrating through the center (point C in Fig.~\ref{fig:cluster}). \textbf{(a)} The densities on sites A (dashed line) and B (full) closest to the projectile. \textbf{(b)} site-averaged double occupation, Eq.~(\ref{eq:d-av}). After Ref.~\cite{balzer_prl_18}.} 
  %%\begin{changed}
  %%Karsten: bitte Bild ersetzen: nur $Z=2$, vertikalen Massstab oben vergoessern. v nur in 1 Bild, es genuegt eine x-Achse - Bilder naeher zusammenschieben. Bezeichnungen (a) und (b) tauschen.
  %%\end{changed}
%  }
  \label{fig:doublons}
  \end{figure} 
is proportional to the correlation energy in the system. In fact, the numerical analysis shows that a charged projectile with an impact energy in the range of a few hundred electron volts may, indeed create a significant number of doublons \cite{balzer_prl_18}. A typical example is shown in Fig.~\ref{fig:doublons} %where the time evolution of the doublon number at two lattice sites B and A adjacent to the impact point is shown 
for a strongly correlated ($U/J=10$) finite graphene cluster. In part (a) we show the electron densities at two lattice sites B and A adjacent to the impact point. During the impact of the projectile ($t=t_0$) electrons from the second nearest site (A) are attracted towards the nearest site B whereas the mean density remains almost constant. After the projectile has left, both densities, with some retardation, return to their initial values. Consider now the associated dynamics of the mean double occupations at sites A and B. While here, too, doublons are transferred from site A to B, the mean value, $d_{\mathrm{av}}$ increases significantly. Most importantly, after the projectile has left,  $d_{\mathrm{av}}$ does not return to its initial value but remains at a significantly larger value. We conclude that the projectile has deposited correlation energy in the system that remains stored there. This is also confirmed by comparison with the uncorrelated average doublon number, Eq.~(\ref{eq:di-hartree}), which follows the average density and, hence, remains almost constant.
In a quantum-mechanical language, under the action of the projectile, the electron system undergoes a transition to an excited state that is associated with a higher doublon occupation \cite{balzer_prl_18}. This explanation is directly confirmed by a representative dimer model that is discussed in Sec.~\ref{ss:dimer}.\\

\subsection{Analytical dimer model.}\label{ss:dimer}
%
%\begin{enumerate}
%\item Modell f\"ur Dimer beschreiben
%\item $W(t)$ zuerst allgemein, Wahl von Gauss motivieren
%\item Fokus auf $W_0$-Abh\"angigkeit, fuer Gauss
%\item zwei Werte U, daf\"ur $W_0$ und v variieren
%\end{enumerate}

\begin{figure*}[h]
\sidecaption
    \centering
    \includegraphics*[width=0.65\linewidth]{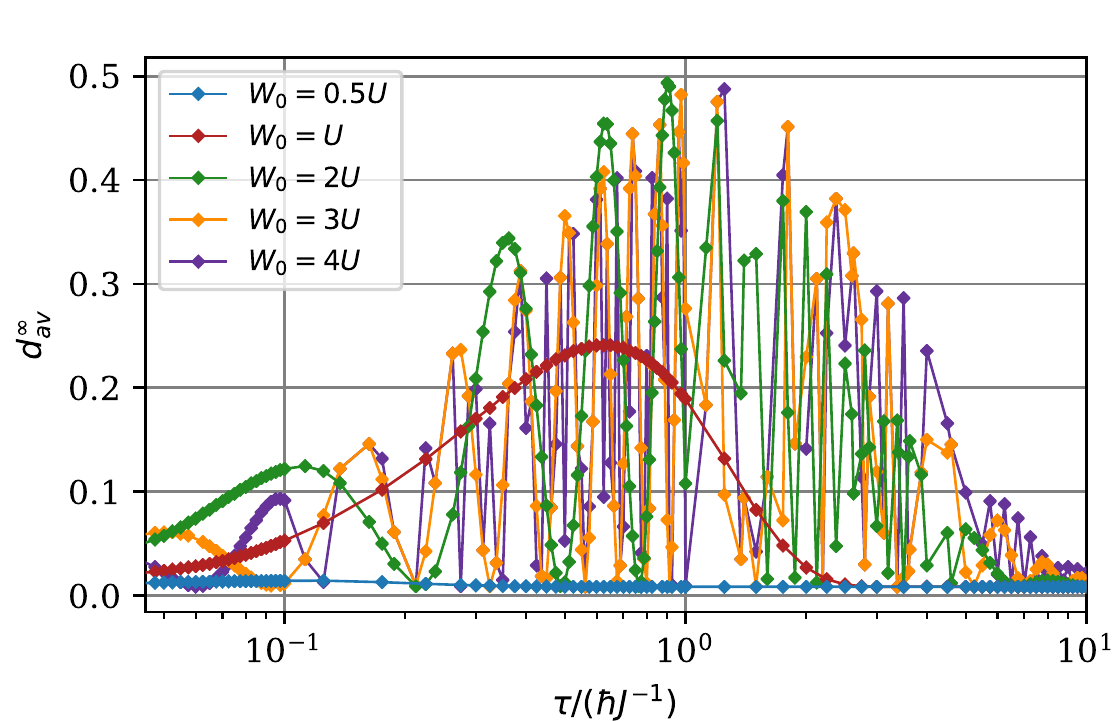}
    \caption{Asymptotic value of the average double occupation, Eq.~(\ref{eq:d-av}), of the dimer versus $\tau$ (proportional to the inverse projectile velocity) for U=15 and different excitation amplitudes, $W_0$. When $W_0$ exceeds $U$ doublons are excited and remain in the system.} 
    \label{fig:dimer}
\end{figure*}
For a qualitative examination of the doublon generation in the system of Fig.~\ref{fig:cluster}, the simplest possible setup is a dimer consisting of only the two sites, A and B, being driven by a pulsed attractive external potential. Since we expect that the excitation of doublons is governed only by the potential difference on sites A and B, it is sufficient to consider the excitation only on one site (B). The time dependence of the excitation is chosen as 
\begin{equation}
W(t)=-W_0\exp^{-(t-t_0)^2/2\tau^2}\,,
\label{eq:gauss-projectile}    
\end{equation}
which closely resembles a positively charged projectile passing close to one site, where the two parameters $W_0$ and $\tau$ have clear implication as the amplitude (proportional to the charge of the ion) and the interaction duration (proportional to one over the velocity), respectively.
For sufficiently large $U$ this can lead to a significant and lasting increase of the mean double occupation $d^\infty_{\mathrm{av}}$, Eq.~(\ref{eq:d-av}). However $d^\infty_{\mathrm{av}}$  strongly depends on $W_0$ and $\tau$, as is confirmed by exact diagonalization results that are shown in Fig. \ref{fig:dimer}. For an excitation amplitude $W_0$ smaller than U, the Hubbard-gap prevents the creation of doublons. For $W_0 > U$ doublon production is possible, and for larger $W_0$,  oscillations caused by transient Bloch oscillations, are observed \cite{balzer_prl_18} the frequency of which grows with $W_0$. Interestingly, the envelopes of these curves are very similar to the stopping-power curves, cf. Fig.~\ref{fig:stopping-curve}. There the total energy gain of the electrons was plotted vs. kinetic energy of the projectile
%where the excitation is similar to a charged particle traveling with 
which here corresponds to the inverse of $\tau^2$. The results of Fig.~\ref{fig:dimer} reflect the fraction of the projectile energy that is transferred  into an increase of the double occupation in the target, and a detailed analysis of the different energy contributions remains to be performed in future work.
The most notable result is, that for an optimal choice of $\tau$ and $W_0$ a permanent increase of the double occupation of up to $0.5$ per site can be achieved, in agreement with the simulation results of Fig.~\ref{fig:doublons}.

We have shown in Ref.~\cite{balzer_prl_18} that the dimer model captures the excitation physics not only qualitatively correctly. Using a Landau-Zener \cite{landau_32,zener_32}  approach the probability for doublon excitation of our model agrees even semi-quantitatively with the simulation results for the $L=12$ cluster of Fig.~\ref{fig:cluster} and shows the correct trends also for other systems, including the optimal coupling strength and projectile velocity that maximize the induced doublon number.\\

\subsection{Doublon dynamics excited by multiple ion impacts.}\label{ss:many-impacts}
\begin{figure}[h]%
\includegraphics*[width=\linewidth]{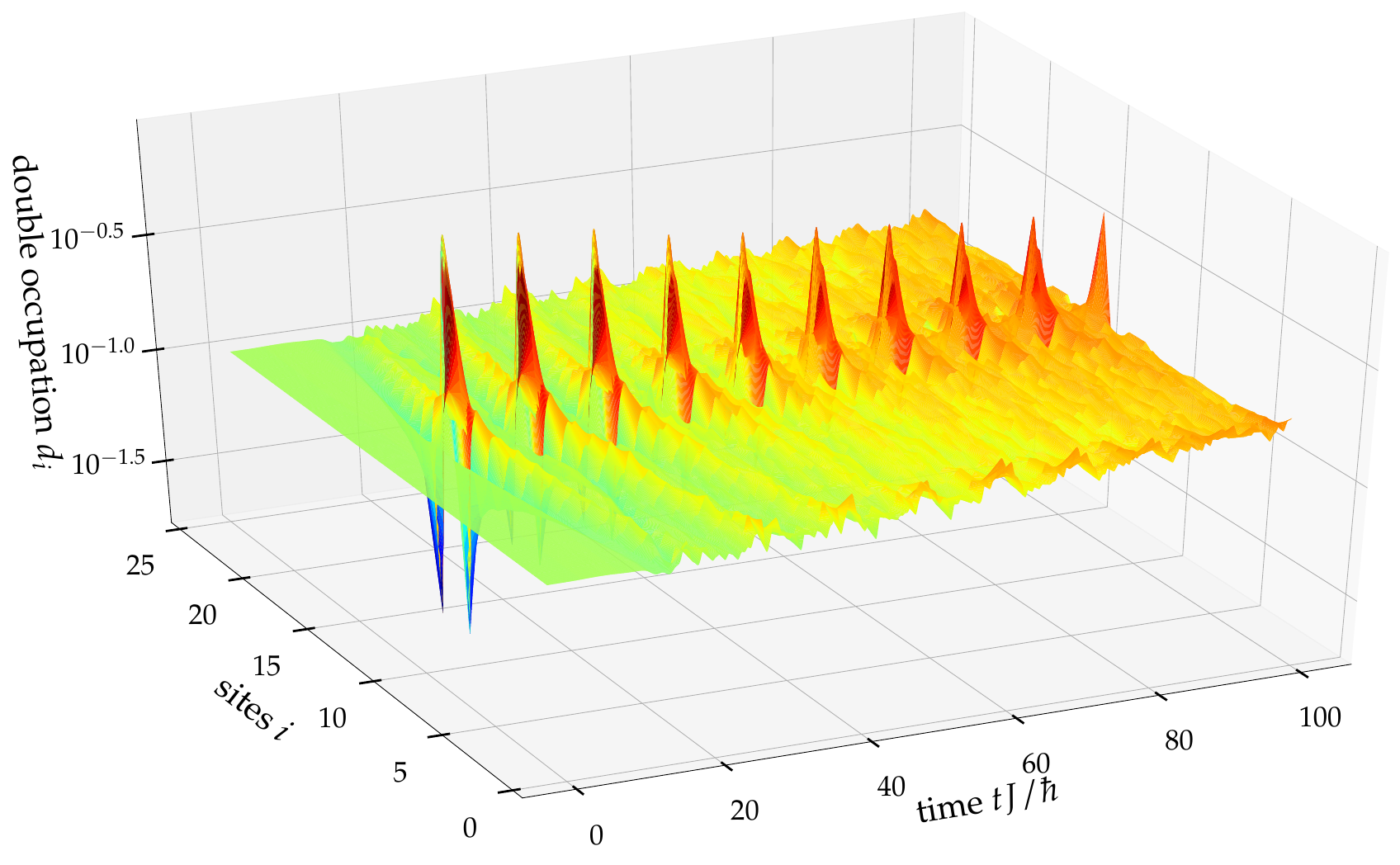}
\caption{%
Time evolution of the site-resolved doublon number, $d_i(t)$, Eq.~(\ref{eq:mean-di}), for a 1D cluster with $L=24$ sites (periodic boundary conditions) and $U/J=4$. The cluster is excited by ten ion impacts in the center (at site 12) using the Gaussian model (\ref{eq:gauss-projectile}). The increase of $d^\infty_{\mathrm{av}}(t)$ can be seen from the slope of the surface. Note the logarithmic scale.
}
\label{fig:stopping-di-multiple}
\end{figure}

The average doublon number in the system can be further increased by repeating the impact once or even more often. The analysis presented in Ref. \cite{balzer_prl_18} showed that this allows to achieve an asymptotic average doublon number of $d_{\mathrm{av}}=0.25$ and even larger. A representative example is shown in Fig.~\ref{fig:stopping-di-multiple}.
At each impact the projectile rapidly increases $d_i$ at the impact point, at the expense of the doublon number at the two nearest neighbor sites. This is followed by a spreading of $d_i(t)$ along the chain (notice the wave fronts). At the same time, with each successive impact, the average doublon number can be systematically increased which can be seen from the increasing doublon level in the foreground.
In that figure the excitation is intentionally kept localized at the same central site in order to monitor the propagation of the doublon occupation along the cluster. Note that, when one uses a Coulomb potential, its long range affects simultaneously many electrons which gives rise to even larger values of $d^\infty_{\mathrm{av}}$ \cite{balzer_prl_18}.

The ion induced nonequilibrium dynamics of the electron system can also be tracked in the spectral function which can be directly measured in photoemission experiment. In Fig.~\ref{fig:stopping-di-multiple} we plot the photoemission spectrum, Eq.~(\ref{eq:photoemission}), that gives information about the occupied energies. The projectiles induce transitions of electrons into the upper Hubbard band corresponding to $\omega>0$. With each successive impact the spectral weight (corresponding to the fraction of electrons) in the upper Hubbard band grows, cf. the shaded areas. 

As in the case of a single impact, Fig.~\ref{fig:doublons}, also after multiple impacts, the many-electron system remains in the excited state characterized by a significantly increased average doublon occupation $d^\infty_{\mathrm{av}}$, after all projectiles have left. This stationary nonequilibrium state will be stable until additional dissipation channels (e.g. to phononic degrees of freedom) set in and is another example of a pre-thermalized state, e.g. \cite{kollar_prb_11,joura_prb_15}. In contrast to previous spatially homogeneous doublon excitation scenarios that used  time-dependent electric fields or a modulation of the lattice depth, e.g. \cite{tokuno_pra_12}, here a local excitation is used that has much more degrees freedom, including timing and locations of the impacts, and a potential to achieve higher doublon numbers and an increased stopping power.
\begin{figure}[h]%
\includegraphics*[width=\linewidth]{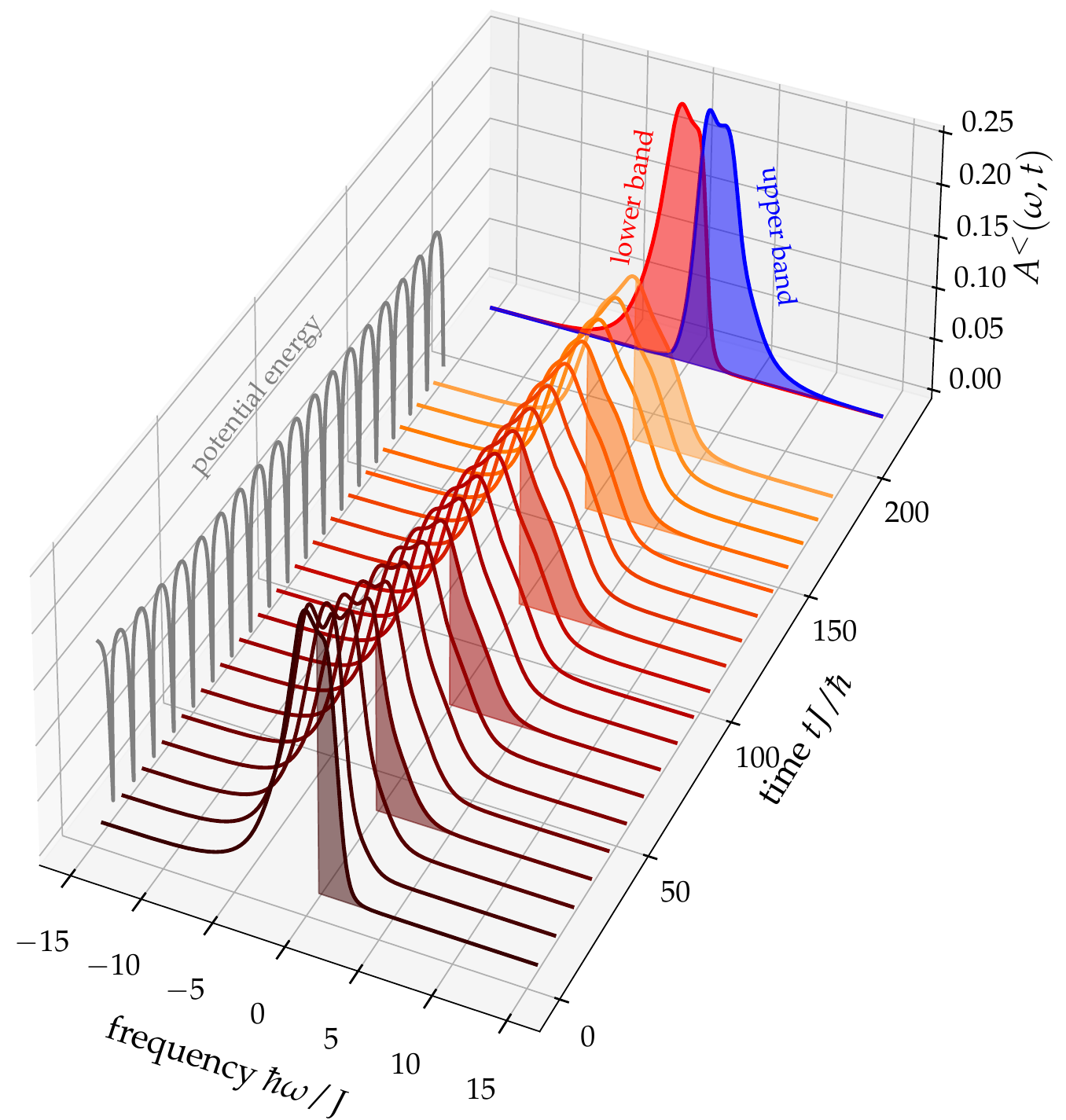}
\caption{%
Time evolution of the spectral function of the occupied states, $A^<$, (photoemission spectrum) for the honeycomb cluster with $L=12$ sites, Fig.~\ref{fig:cluster}, and $U/J=4$, from a two-time NEGF simulation with second order Born selfenergies. The cluster is excited by twenty equidistant ion impacts (at times $10, 20, \dots 200$) in the center using a Coulomb potential for the electron-projectile interaction, cf. Eq.~(\ref{eq:ham1}) and the grey line on the left. The projectiles lead to an increased occupation of the upper Hubbard band corresponding to $\omega > 0$, cf. the shaded areas. The spectra are shown at times $5, 15, \dots 195$ with a width of the probe pulse [cf. Eq.~(\ref{eq:photoemission})] of $\kappa=2.5\hbar/J$ which causes a spectral broadening. For comparison, 
we also present the results of an unexcited cluster, where only the lower Hubbard band is occupied, depicted by the red filled area corresponding to $A^<$, whereas the unoccupied upper Hubbard band ($A^>$) is shown by the blue area.
}
\label{fig:stopping-a-multiple}
\end{figure}

\section{Embedding scheme to capture charge transfer dynamics between projectile and target.}\label{s:embedding}

So far we have considered only the case of high projectile velocities where the feedback from the surface to the ion is small and restricted to a reduction of its velocity whereas quantum effects are neglected. On the other hand, when the impact velocity is reduced, the interaction duration of the projectile with the lattice increases and electron transfer between both systems may occur.

Quantum transitions inside the projectile and charge transfer have been studied approximately with quantum kinetic models (Newns-Anderson model) where the projectile was treated as a few level system \cite{pamperin_2015_many}. Furthermore, there have been a number of TDDFT studies of ions impinging onto correlated materials such as graphene or boron nitride (BN) \cite{zhao_2015_comparison,ojanpera_prb_14} and on finite systems such as metal clusters \cite{Moss_2009_Li+_AlCluster} \cite{Castro2012_H+Li4},    carbon nanostructures \cite{Krasheninnikov2007}, or
graphene fragments \cite{Bubin2012} (for more references see \cite{bonitz_fcse_18}), where quantum transitions inside the projectile are taken into account.
However, the uncertainties in the quality of the adiabatic LDA and the model parameters in the Newns-Anderson model, respectively, as well as the neglect of correlation effects in the material \cite{bonitz_fcse_18} make it desirable to develop an independent many-body approach to this problem. 

Here, we present a nonequilibrium Green functions approach for the electron transfer dynamics between projectile and a strongly correlated solid. We start from the second-quantized many-body Hamiltonian for the electrons in the plasma-solid interface and separate the system into a plasma ($p$) and solid surface part ($s$) [we denote $\Omega=\{p,s\}$ and do not write the spin index explicitly],
\begin{eqnarray}
\label{eq.ham}
H_{\textup{interface}} &=&\sum_{\alpha\beta\in\Omega}\sum_{ij}H^{\alpha\beta}_{ij}(t)c^{\alpha\dagger}_ic^\beta_j+
\nonumber\\
&&\frac{1}{2}\sum_{\alpha\beta\gamma\delta\in\Omega}\sum_{ijkl}W^{\alpha\beta\gamma\delta}_{ijkl}c^{\alpha\dagger}_ic^{\beta\dagger}_jc^{\gamma}_kc^{\delta}_l\,.
\end{eqnarray}
Here, the operator $c^{\alpha\dagger}_i$ ($c^{\alpha}_i$) creates (annihilates) an electron in the state $i$ of part $\alpha$. The one-particle Hamiltonian~$H(t)$ contains the kinetic and the time-dependent potential energy of electrons, and $W$ accounts for all possible electron-electron Coulomb interactions within and between the two parts.

Considering individual energetic plasma ions, which penetrate into the solid, undergo scattering and stopping in the surface layers or are reflected, we describe the system~(\ref{eq.ham}) by a one-particle nonequilibrium Green function (\ref{eq:negf}), $G^{\alpha\beta}_{ij}(t,t')$, which now has an additional $2\times2$ matrix structure ($\alpha, \beta=\{p,s\}$), 
\begin{eqnarray}
\label{eq.negf}
 G^{\alpha\beta}_{ij}(t,t') &=&- \mathrm{i}\hbar \langle T_C c^\alpha_{i}(t)c_{j}^{\beta\dagger}(t')\rangle\,, 
 \\
  \rho_{ij}^{\alpha\beta}(t) &=& - \mathrm{i}\hbar G^{\beta\alpha}_{ji}(t,t^+)\,,
\label{eq:g-dm}  
\end{eqnarray}
e.g., Refs.~\cite{stefanucci_cambridge_2013,balzer-book}, and the time-diagonal elements provide the density matrix (\ref{eq:g-dm}).
The diagonal elements, $\rho_{ij}^{pp}$ [$\rho_{ij}^{ss}$], refer to the plasma part, describing the dynamics of free electrons and electrons bound in the ion [to the solid part,  describing electrons in bound states of the solid surface]. Moreover, the density matrix component $\rho_{ij}^{ps}$ is related to charge transfer processes between plasma and solid and will be of special interest in the following. 

The equations of motion for the NEGF are the generalization of Eq.~(\ref{eq.kbe})
to the plasma-solid interface, 
%($\hbar=1$):
%
\begin{eqnarray}\label{eq.kbe_interface}
 \mathrm{i}\hbar\,\partial_tG^{\alpha\beta}_{ij}(t,t') &-& \sum_{\delta\in\Omega,k} H^{\alpha\delta}_{ik}(t)G^{\delta\beta}_{kj}(t,t')=
 \\ \nonumber
 \delta^{\alpha\beta}_{ij}\delta_C(t,t') &+
 &\sum_{\delta\in\Omega,k} \int_C\!\!\!d\bar{t}\,\Sigma^{\alpha\delta}_{ik}[W,G](t,\bar{t})G^{\delta\beta}_{kj}(\bar{t},t')\,,
\end{eqnarray}
where the self-energy $\Sigma^{\alpha\beta}(t,t')$ describes the interaction between the electrons and with phonons.
%and depends, in general, on all components of $W$ and  %$G(t,t')$. 
Even though a complete solution of the KBE~(\ref{eq.kbe_interface}) for real materials and with a full quantum treatment of the plasma electrons is out of reach, these equations provide the rigorous starting point for the development of consistent approximations. 
In the following we show how it is possible to include the electronic states of the ion via an embedding self-energy approach~\cite{stefanucci_cambridge_2013}, where resonant (neutralization and ionization) processes can be studied. While this embedding approach is based on a formal decoupling of the surface and plasma parts of the KBE, it retains one-electron charge transfer in the Hamiltonian $H^{sp}$, cf. Eq.~(\ref{eq.sigma.ct}), see below. 
%This represents a particularly feasible scheme that is explained below.
A closed description of the solid can be maintained if correlations in the plasma part and the feedback of the solid on the plasma can be neglected, i.e., for $\Sigma^{sp}\approx\Sigma^{pp}\approx 0$. This is usually well fulfilled in plasmas, except for  plasmas at or beyond atmospheric pressure or in warm dense matter \cite{DORNHEIM_physrep18} where small correlation corrections should be taken into account.  Then, the KBE~(\ref{eq.kbe}) for the plasma part simplify to
\begin{eqnarray}
  \sum_{k} \left\{\mathrm{i}\hbar\partial_t\delta_{ik}-H^{pp}_{ik}(t)\right\}g^{pp}_{kj}(t,t')=\delta_{ij}\delta_C(t,t')\,,
\end{eqnarray}
 where the solution $g^{pp}(t,t')$ denotes the NEGF of the electrons inside the plasma ions [here we do not consider processes involving free electrons in the plasma phase because the do not contribute to charge transfer except for heavy particle induced secondary electron emission], whereas the time dependence of $H^{pp}(t)$ accounts for possible parametric changes of the energy levels (e.g., as function of the distance of the ion from the surface). 
 
 The main result of the embedding procedure is a closed equation for $G^{ss}(t,t')$:
\begin{eqnarray}
\label{eq.kbe.embedding}
 \sum_{k} \left\{\mathrm{i}\hbar\partial_t\delta_{ik}-H^{ss}_{ik}(t)\right\}G^{ss}_{kj}(t,t')=\delta_{ij}\delta_C(t,t')
 +
 \\\nonumber
 \sum_k\int_C\!\!\!d\bar{t}\,\left\{\Sigma^{ct}_{ik}(t,\bar{t})+\Sigma^{ss}_{ik}[G^{ss}](t,\bar{t})\right\}G^{ss}_{kj}(\bar{t},t')\,,
\end{eqnarray}
to be complemented with the adjoint equation, with the charge transfer (or embedding) self-energy that involves the charge transfer hamiltonian
\begin{eqnarray}
\label{eq.sigma.ct}
  \Sigma^{ct}_{ij}(t,t') &=& \sum_{kl}H^{sp}_{ik}(t)g_{kl}^{pp}(t,t') H^{ps}_{lj}(t')\,,
\\
H^{sp}_{ij}(t) &=& \int\!\!\! d^3r\,\phi^s_i(\vec{r})(\hat{T}+\hat{V})\phi_j^p(\vec{r};t)\,.
\label{eq:hsp}
\end{eqnarray}
Equation~(\ref{eq.kbe.embedding}) shows how the many-body description of an isolated (but correlated) solid is altered by the presence of the electronic states of a plasma ion (or neutral), with the latter giving rise to an additional self-energy $\Sigma^{\textup{ct}}(t,t')$. While,  for $\Sigma^{\textup{ct}}=0$, the KBE~(\ref{eq.kbe.embedding}) conserve the particle number and total energy [for a conserving approximation of the self-energy $\Sigma^{ss}$, such as the ones discussed in Sec.~\ref{ss:sigmas}], the inclusion of the embedding self-energy explicitly allows for time-dependent changes of the particle number (and energy) in the solid and, thus, accounts for ion charging and neutralization effects. For the practical solution of Eq.~(\ref{eq.kbe.embedding}), the charge transfer Hamiltonian $H^{sp}(t)$ has to be computed by selecting the relevant electronic transitions between solid and plasma and computing the matrix elements of the kinetic and potential energy operators $\hat{T}$ and $\hat{V}$, with the electronic single-particle wave functions $\phi^{s}$ ($\phi^p$) in the solid (ion).
  \begin{figure}[h]
  \begin{center} 
  \includegraphics[width=0.325\textwidth]{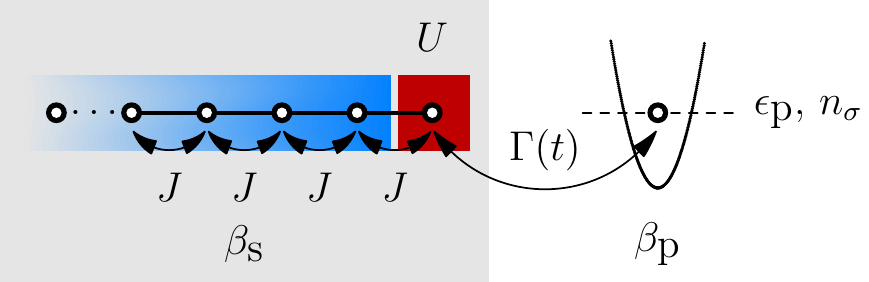}\hspace*{1.0cm}\\[0.25cm]
  \parbox{0.255\textwidth}{
  \includegraphics[width=0.255\textwidth]{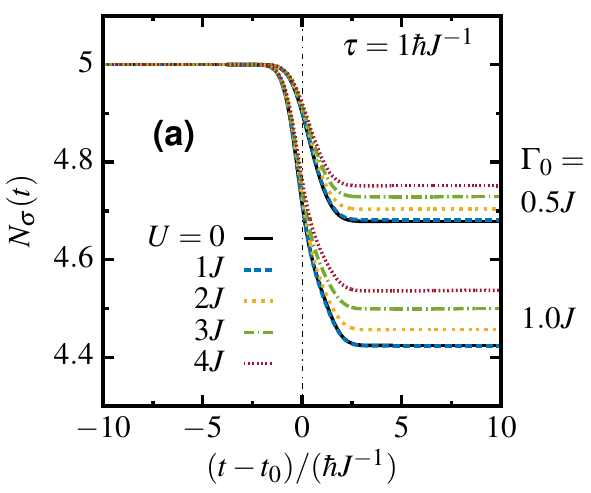}
  \includegraphics[width=0.255\textwidth]{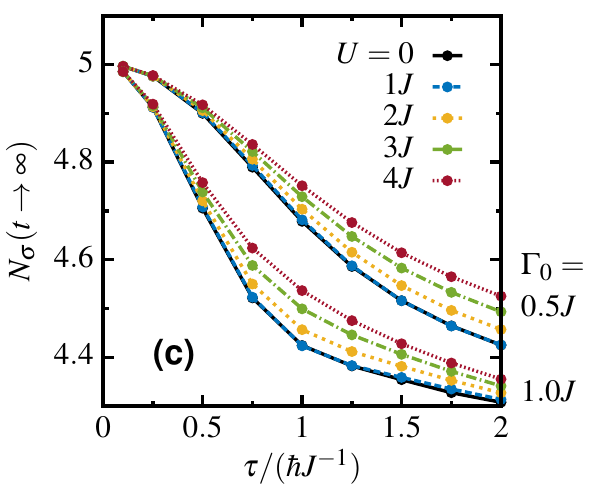}
  }
  \parbox{0.2175\textwidth}{
  \includegraphics[width=0.2175\textwidth]{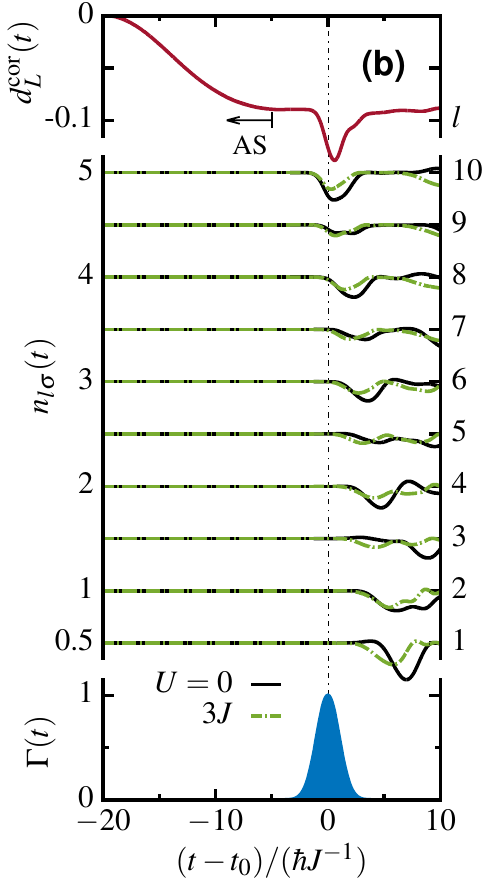}
  }
  %
  %%\hspace{-0.cm}\includegraphics[width=0.25\textwidth]{figures/embedding.pdf}\hspace{0.1cm}
  %%\hspace{-0.cm}\includegraphics[width=0.22\textwidth]{figures/embedding1.pdf}\\[0.1cm]
  %%\hspace{-0.cm}\includegraphics[width=0.485\textwidth]{figures/embedding2.pdf}
  \end{center}
  \vspace{-.40cm}
  \caption{
  Numerical illustration of the embedding scheme. 
  An initially half-filled tight-binding chain ($L=10$ sites, nearest-neighbor hopping $J$, Hubbard interaction strength $U$ on the last site, and inverse temperature $\beta_s=100J^{-1}$) is interacting with an external energy level $\epsilon_\textup{p}=J$ via a time-dependent coupling $\Gamma(t)=\Gamma_0\textup{e}^{-(t-t_0)^2/(2\tau^2)}$, cf. bottom of part (b), giving rise to the transfer of charge. The initial occupation of the energy level is given by $n_\sigma=0.269$ (corresponding to a Fermi distribution with an inverse temperature $\beta_\textup{p}=1J^{-1}$). 
  \textbf{(a)} Time evolution of the total particle number $N_\sigma(t)$ for different $U$ and $\Gamma_0$, computed from Eq.~(\ref{eq.kbe.embedding}) with a local second Born self-energy ($\Sigma_{ij}=\delta_{ij}\Sigma_i$); $\tau=1\hbar J^{-1}$. \textbf{(b)} Local electron densities $n_{l\sigma}(t)$ and correlation part of the double occupation $d_{L}^{\textup{cor}}(t)=d_L(t)-d^{\textup{H}}_L(t)$ on the last site $l=L=10$ for $U=3$, $\Gamma_0=1J$ and $\tau=1\hbar J^{-1}$; for times $t-t_0\lesssim-5$ the time evolution of $d_L^{\textup{cor}}(t)$ corresponds to the ground state preparation by adiabatic switching (AS). \textbf{(c)} Final values of the particle number $N_\sigma(t\rightarrow\infty)$ as function of the interaction time $\tau$ for different $U$ and $\Gamma_0$.
  %%
  %%Numerical illustration of the embedding scheme. 
  %%An initially half-filled tight-binding chain ($20$ sites, nearest-neighbor hopping $J$, coupling strength $U$ on the first site, and inverse temperature $\beta_\textup{s}=100J^{-1}$) is interacting with an approaching classical projectile via the time-dependent potential (\ref{eq:gauss-projectile}). %$\Gamma(t)=\Gamma_0\textup{e}^{-(t-t_0)^2/(2\tau^2)}$ 
  %%The electrons are coupled to an external energy level $\epsilon_\textup{i}=+J$ giving rise to the transfer of charge. The initial occupation of the energy level is given by $n_\sigma=0.269$ (corresponding to an inverse temperature of $\beta_\textup{i}=1J^{-1}$) and $\bar{n}_\sigma=1-n_\sigma$. Furthermore, $N_\sigma(t)=\sum_{i}n_{i\sigma}(t)$ denotes the total electron density on the chain.
  }
  
%%\begin{changed}
%%Karsten: eine Abb. aehnlich zum review, mit korrelierter letzter site.  Wenn moeglich, Abhaengigkeit von U und von Geschwindigkeit demonstrieren. Beim Zeitverhalten habe ich auf das Gauss-Modell verwiesen, Amplitude ist $W_0$. Ist eine kontinuierliche Aenderung von $W_0$ sinnvoll (Ladung)? Ist die Angabe der Temperatur $\beta_i$ sinnvoll? Die Besetzung haengt doch genauso vom Energiespektrum (Normierung) ab.
%%\end{changed}
  \label{fig:embedding}
  \end{figure} 

A first test of this embedding scheme is shown in Fig.~\ref{fig:embedding}, where a correlated Hubbard chain (for simplicity only the last site is correlated) is coupled to a single active energy level $\epsilon_p = J$ of an approaching ion via the charge transfer hamiltonian 
$H^{sp}_{i}(t)= \delta_{iL}\Gamma(t)$,
cf. the sketch on top of Fig.~\ref{fig:embedding}. The time dependence of $H^{sp}_i$ is approximated by $\Gamma(t)=\Gamma_0\textup{e}^{-(t-t_0)^2/(2\tau^2)}$, and the initial occupation of the energy level $\epsilon_p$ is set to $n_\sigma=0.269$.

The charge transfer from the chain to the ion, seen in the reduction of the total electron number in the chain, $N_\sigma(t)=\sum_i n_{i\sigma}(t)$, is shown as a function of time in Fig.~\ref{fig:embedding}.(a). The reduction of $N_\sigma$ is found to be nearly proportional to the ion charge (amplitude $\Gamma_0$) up to the resonance condition $\Gamma_0=J$. Thus, as expected, a highly charged ion will be more strongly neutralized. For $\Gamma_0>J$, away from resonance, the net transfer of charge will decrease again. The neutralization time is given by the interaction duration $\tau$ which is inversely proportional to the projectile velocity. The dependence of the magnitude of the charge transfer on $\tau$ is analyzed in Fig.~\ref{fig:embedding}.(c) and again confirms the expected trend: the charge transfer increases with $\tau$, i.e. is larger for slower projectiles, whereas for $\tau \lesssim 0.1 \hbar/J$ it is negligible. %Using parameters typically for graphene -- $a\sim 1\AA, J\sim 1eV$, the value $\tau = 0.1 \hbar/J$ translates into projectile kinetic energies of ....eV, for protons and ....eV for alpha-particles.
%\begin{changed}
%  check the data for the energies! Diese Abschaetzung ist schwierig, vielleicht eher weglassen: Fuer $v=a/\tau=1J/(0.1\hbar)$\,\AA  wuerde man fuer Protonen etwa $120$ keV bekommen.
%\end{changed}
Figure~\ref{fig:embedding} (b) shows the spatial propagation of the removed charge (hole) along the chain as a function of time (the distortion of the dip is due to reflections from the edge of the chain). Again one sees that, in the presence of correlations, the propagation speed is reduced, in agreement with simulations of fermion propagation in optical lattices \cite{schluenzen_prb16,schluenzen_cpp16}.

Finally, we can analyze the effect of correlations in the target on the charge transfer. As can be seen in Fig.~\ref{fig:embedding}.(a) and (c), an increase of electron-electron correlations reduces the charge transfer, which is a consequence of the reduced mobility of the electrons in the chain. An increase of the interaction strength from zero to $U=4/J$, which is a realistic range for graphene nanoribbons, reduces the charge transfer by about 20$\%$, in the present setup. 

In conclusion, we have demonstrated a NEGF approach to charge transfer between a plasma ion and a strongly correlated finite electron system. The next task is to derive improved data for the energy levels and occupations of the projectile. Further, the resonant charge transfer, studied in this section, and the energy deposition and electronic excitation of the target that were discussed in Sec.~\ref{s:results}, should be integrated into a single model to take into account the mutual influences of both processes.
%However, we note that a more adequate model for resonant ion-target interactions should also include the one-particle potential of Eqs.~(\ref{eq:ham1}) or (\ref{eq:gauss-projectile}), which will modify the observed trends.

\section{Summary and Discussion.}\label{s:discussion}
In this paper we studied strongly correlated inhomogeneous finite systems of fermions such as electrons in graphene clusters and nanoribbons. We considered the electronic response to a spatially and temporally localized excitation by a charged particle. Using a nonequilibrium Green functions (NEGF) approach we computed the time-dependent interaction of the projectile with the many-electron system and the dependence of the energy transfer on the impact energy \cite{balzer_prb16}. An interesting observation was that, a low projectile energies, correlation effects lead to enhanced energy transfer. Our analysis revealed that the ion impact 
 causes a transition of the system across the Hubbard gap leading to the formation of doubly occupied lattice sites (doublons) \cite{balzer_prl_18}. We investigated the spatial propagation of the doublon number across the cluster. Eventually a homogeneous nonequilibrium steady state is reached that is long lived and may have interesting electronic and optical properties.
 A physically intuitive picture was given in terms of an analytical model for a two-site system where the doublon formation is explained in terms of a two-fold passage of an avoided crossing (Landau-Zener picture \cite{balzer_prl_18}).
The effect should be particularly important for strongly correlated finite systems, such as graphene nanoribbons.
%in contact with a high-pressure plasma 
For an experimental observation the best candidates are fermionic atoms in optical lattices. There the projectile impact can be easily mimicked by a proper time-dependent modulation of the lattice potentials nearest to the ``impact'' point.

We demonstrated that doublon formation and propagation in correlated finite lattice systems can be accurately simulated with NEGF. In addition to two-time results we presented single-time results within the generalized Kadanoff-Baym ansatz (GKBA) with Hartree-Fock propagators (HF-GKBA). To further improve the accuracy of GKBA calculations in the future, we introduced the correlated GKBA (C-GKBA) that allows to systematically incorporate correlation effects in the propagators $G^{R/A}$. Moreover, we discussed how to systematically take into account initial correlations in the GKBA and presented an idea that is complementary to recent results for equilibrium correlations \cite{karlsson_gkba18,verdozzi_gkba-note}. 

Aside from an accurate treatment of correlation effects, quantitatively reliable NEGF results also require to improve the underlying model. One way to go beyond the present one-band Hubbard model is to use an extended Hubbard model as demonstrated in Ref.~\cite{joost_pss_18}, or to perform \textit{ab initio NEGF} simulations using a Kohn-Sham basis, e.g. on the basis of the Yambo code \cite{marini_2009_yambo}.

\begin{acknowledgement}
We acknowledge a grant for computing time at the HLRN.
\end{acknowledgement}

% Use the following code if you wish to generate your bibliography with BibTeX;
% replace the string "pss_demo" below with the name(s) of
% the BibTeX data base(s) you want to use.
% The resulting bibliography-output (the content of the .bbl file)
% must be pasted back into this file before submission.
% Please also include your BibTeX data base file(s) in your submission
% so that we can re-run BibTeX if necessary.
%
\bibliographystyle{pss}
%\bibliography{pss_demo,mb-ref,dfg_pngf,dft}

\begin{thebibliography}{[10]}

\bibitem{adamovich_2017_plasma}% article
 \textsc{I.~Adamovich},  \textsc{S.\,D. Baalrud},  \textsc{A.~Bogaerts},
  \textsc{P.\,J. Bruggeman},  \textsc{M.~Cappelli},  \textsc{V.~Colombo},
  \textsc{U.~Czarnetzki},  \textsc{U.~Ebert},  \textsc{J.\,G. Eden},
  \textsc{P.~Favia},  \textsc{D.\,B. Graves},  \textsc{S.~Hamaguchi},
  \textsc{G.~Hieftje},  \textsc{M.~Hori},  \textsc{I.\,D. Kaganovich},
  \textsc{U.~Kortshagen},  \textsc{M.\,J. Kushner},  \textsc{N.\,J. Mason},
  \textsc{S.~Mazouffre},  \textsc{S.\,M. Thagard},  \textsc{H.\,R. Metelmann},
  \textsc{A.~Mizuno},  \textsc{E.~Moreau},  \textsc{A.\,B. Murphy},
  \textsc{B.\,A. Niemira},  \textsc{G.\,S. Oehrlein},  \textsc{Z.\,L.
  Petrovic},  \textsc{L.\,C. Pitchford},  \textsc{Y.\,K. Pu},
  \textsc{S.~Rauf},  \textsc{O.~Sakai},  \textsc{S.~Samukawa},
  \textsc{S.~Starikovskaia},  \textsc{J.~Tennyson},  \textsc{K.~Terashima},
  \textsc{M.\,M. Turner},  \textsc{M.\,C.\,M. van\,de Sanden},  and
  \textsc{A.~Vardelle}\iffalse The 2017 plasma roadmap: Low temperature plasma
  science and technology\fi,
 \jr{J. Phys. D: Appl. Phys.} \textbf{50}(32), 323001 (2017).


\bibitem{bonitz_fcse_18}% article
 \textsc{M.~Bonitz},  \textsc{V.~Filinov},  \textsc{J.~Abraham},
  \textsc{K.~Balzer},  \textsc{H.~Kaehlert},  \textsc{E.~Pehlke},
  \textsc{F.\,X. Bronold},  \textsc{M.~Pamperin},  \textsc{M.~Becker},
  \textsc{D.~Loffhagen},  and  \textsc{H.~Fehske}\iffalse Towards an integrated
  modeling of the plasma-solid interface\fi,
 \jr{arXiv:} (2018),
submitted for publication.


\othercit
\bibitem{sigmund_springer_06}% book
 \textsc{P.~Sigmund},
Particle Penetration and Radiation Effects (Springer, 2006).


\bibitem{nagy_pra_98}% article
 \textsc{I.~Nagy} and  \textsc{B.~Apagyi}\iffalse Scattering-theory formulation
  of stopping powers of a solid target for protons and antiprotons with
  velocity-dependent screening\fi,
 \jr{Phys. Rev. A} \textbf{58}(Sep), R1653--R1656 (1998).


\bibitem{pitarke_prb_95}% article
 \textsc{J.\,M. Pitarke},  \textsc{R.\,H. Ritchie},  and  \textsc{P.\,M.
  Echenique}\iffalse Quadratic response theory of the energy loss of charged
  particles in an electron gas\fi,
 \jr{Phys. Rev. B} \textbf{52}(Nov), 13883--13902 (1995).


\bibitem{quijada_pra_07}% article
 \textsc{M.~Quijada},  \textsc{A.\,G. Borisov},  \textsc{I.~Nagy},
  \textsc{R.\,D. Mui\~no},  and  \textsc{P.\,M. Echenique}\iffalse
  Time-dependent density-functional calculation of the stopping power for
  protons and antiprotons in metals\fi,
 \jr{Phys. Rev. A} \textbf{75}(Apr), 042902 (2007).


\bibitem{ojanpera_prb_14}% article
 \textsc{A.~Ojanper\"a},  \textsc{A.\,V. Krasheninnikov},  and
  \textsc{M.~Puska}\iffalse Electronic stopping power from first-principles
  calculations with account for core electron excitations and projectile
  ionization\fi,
 \jr{Phys. Rev. B} \textbf{89}(Jan), 035120 (2014).


\bibitem{zhao_2015_comparison}% article
 \textsc{S.~Zhao},  \textsc{W.~Kang},  \textsc{J.~Xue},  \textsc{X.~Zhang},
  and  \textsc{P.~Zhang}\iffalse Comparison of electronic energy loss in
  graphene and {BN} sheet by means of time-dependent density functional
  theory\fi,
 \jr{J. Phys. Condens. Matter} \textbf{27}, 025401 (2015).


\bibitem{srim_10}% article
 \textsc{J.\,F. Ziegler},  \textsc{M.~Ziegler},  and
  \textsc{J.~Biersack}\iffalse Srim – the stopping and range of ions in
  matter (2010)\fi,
 \jr{Nuclear Instruments and Methods in Physics Research Section B: Beam
  Interactions with Materials and Atoms} \textbf{268}(11), 1818 -- 1823 (2010),
19th International Conference on Ion Beam Analysis.


\bibitem{Yang2007}% article
 \textsc{L.~Yang},  \textsc{C.\,H. Park},  \textsc{Y.\,W. Son},  \textsc{M.\,L.
  Cohen},  and  \textsc{S.\,G. Louie}\iffalse Quasiparticle energies and band
  gaps in graphene nanoribbons\fi,
 \jr{Phys. Rev. Lett.} \textbf{99}, 186801 (2007).


\bibitem{Han2007}% article
 \textsc{M.\,Y. Han},  \textsc{B.~\"Ozyilmaz},  \textsc{Y.~Zhang},  and
  \textsc{P.~Kim}\iffalse Energy band-gap engineering of graphene
  nanoribbons\fi,
 \jr{Phys. Rev. Lett.} \textbf{98}, 206805 (2007).


\bibitem{nakada_edge_1996}% article
 \textsc{K.~Nakada},  \textsc{M.~Fujita},  \textsc{G.~Dresselhaus},  and
  \textsc{M.\,S. Dresselhaus}\iffalse Edge state in graphene ribbons:
  {Nanometer} size effect and edge shape dependence\fi,
 \jr{Physical Review B} \textbf{54}, 17954--17961 (1996).


\bibitem{son_energy_2006}% article
 \textsc{Y.\,W. Son},  \textsc{M.\,L. Cohen},  and  \textsc{S.\,G.
  Louie}\iffalse Energy {Gaps} in {Graphene} {Nanoribbons}\fi,
 \jr{Physical Review Letters} \textbf{97}, 216803 (2006).


\bibitem{yang_quasiparticle_2007}% article
 \textsc{L.~Yang},  \textsc{C.\,H. Park},  \textsc{Y.\,W. Son},  \textsc{M.\,L.
  Cohen},  and  \textsc{S.\,G. Louie}\iffalse Quasiparticle {Energies} and
  {Band} {Gaps} in {Graphene} {Nanoribbons}\fi,
 \jr{Physical Review Letters} \textbf{99}, 186801 (2007).


\bibitem{ruffieux_electronic_2012}% article
 \textsc{P.~Ruffieux},  \textsc{J.~Cai},  \textsc{N.\,C. Plumb},
  \textsc{L.~Patthey},  \textsc{D.~Prezzi},  \textsc{A.~Ferretti},
  \textsc{E.~Molinari},  \textsc{X.~Feng},  \textsc{K.~M{\"u}llen},
  \textsc{C.\,A. Pignedoli},  and  \textsc{R.~Fasel}\iffalse Electronic
  {Structure} of {Atomically} {Precise} {Graphene} {Nanoribbons}\fi,
 \jr{ACS Nano} \textbf{6}, 6930--6935 (2012).


\bibitem{sode_electronic_2015}% article
 \textsc{H.~S{\"o}de},  \textsc{L.~Talirz},  \textsc{O.~Gr{\"o}ning},
  \textsc{C.\,A. Pignedoli},  \textsc{R.~Berger},  \textsc{X.~Feng},
  \textsc{K.~Müllen},  \textsc{R.~Fasel},  and  \textsc{P.~Ruffieux}\iffalse
  Electronic band dispersion of graphene nanoribbons via {Fourier}-transformed
  scanning tunneling spectroscopy\fi,
 \jr{Physical Review B} \textbf{91}, 045429 (2015).


\bibitem{wang_giant_2016}% article
 \textsc{S.~Wang},  \textsc{L.~Talirz},  \textsc{C.\,A. Pignedoli},
  \textsc{X.~Feng},  \textsc{K.~M{\"u}llen},  \textsc{R.~Fasel},  and
  \textsc{P.~Ruffieux}\iffalse Giant edge state splitting at atomically precise
  graphene zigzag edges\fi,
 \jr{Nature Communications} \textbf{7}, 11507 (2016).


\bibitem{Bai2009}% article
 \textsc{J.~Bai},  \textsc{X.~Duan},  and  \textsc{Y.~Huang}\iffalse Rational
  fabrication of graphene nanoribbons using a nanowire etch mask\fi,
 \jr{Nano Letters} \textbf{9}, 2083--2087 (2009).


\bibitem{Jiao2010}% article
 \textsc{L.~Jiao},  \textsc{X.~Wang},  \textsc{G.~Diankov},  \textsc{H.~Wang},
  and  \textsc{H.~Dai}\iffalse Facile synthesis of high-quality graphene
  nanoribbons\fi,
 \jr{Nature Nanotechnology} \textbf{5}, 321 (2010).


\bibitem{kimouche_ultra_narrow_2015}% article
 \textsc{A.~Kimouche},  \textsc{M.\,M. Ervasti},  \textsc{R.~Drost},
  \textsc{S.~Halonen},  \textsc{A.~Harju},  \textsc{P.\,M. Joensuu},
  \textsc{J.~Sainio},  and  \textsc{P.~Liljeroth}\iffalse Ultra-narrow metallic
  armchair graphene nanoribbons\fi,
 \jr{Nature Communications} \textbf{6}, 10177 (2015).


\bibitem{jensen_ultrafast_2013}% article
 \textsc{S.\,A. Jensen},  \textsc{R.~Ulbricht},  \textsc{A.~Narita},
  \textsc{X.~Feng},  \textsc{K.~M{\"u}llen},  \textsc{T.~Hertel},
  \textsc{D.~Turchinovich},  and  \textsc{M.~Bonn}\iffalse Ultrafast
  {Photoconductivity} of {Graphene} {Nanoribbons} and {Carbon} {Nanotubes}\fi,
 \jr{Nano Letters} \textbf{13}, 5925--5930 (2013).


\bibitem{gierz_tracking_2015}% article
 \textsc{I.~Gierz},  \textsc{F.~Calegari},  \textsc{S.~Aeschlimann},
  \textsc{M.~Chávez~Cervantes},  \textsc{C.~Cacho},  \textsc{R.\,T. Chapman},
  \textsc{E.~Springate},  \textsc{S.~Link},  \textsc{U.~Starke},
  \textsc{C.\,R. Ast},  and  \textsc{A.~Cavalleri}\iffalse Tracking {Primary}
  {Thermalization} {Events} in {Graphene} with {Photoemission} at {Extreme}
  {Time} {Scales}\fi,
 \jr{Physical Review Letters} \textbf{115}, 086803 (2015).


\bibitem{senkovskiy_semiconductor_metal_2017}% article
 \textsc{B.\,V. Senkovskiy},  \textsc{A.\,V. Fedorov},  \textsc{D.~Haberer},
  \textsc{M.~Farjam},  \textsc{K.\,A. Simonov},  \textsc{A.\,B. Preobrajenski},
   \textsc{N.~Mårtensson},  \textsc{N.~Atodiresei},  \textsc{V.~Caciuc},
  \textsc{S.~Blügel},  \textsc{A.~Rosch},  \textsc{N.\,I. Verbitskiy},
  \textsc{M.~Hell},  \textsc{D.\,V. Evtushinsky},  \textsc{R.~German},
  \textsc{T.~Marangoni},  \textsc{P.\,H.\,M. van Loosdrecht},  \textsc{F.\,R.
  Fischer},  and  \textsc{A.~Grüneis}\iffalse Semiconductor-to-{Metal}
  {Transition} and {Quasiparticle} {Renormalization} in {Doped} {Graphene}
  {Nanoribbons}\fi,
 \jr{Advanced Electronic Materials} \textbf{3} (2017).


\bibitem{ivanov_role_2017}% article
 \textsc{I.~Ivanov},  \textsc{Y.~Hu},  \textsc{S.~Osella},  \textsc{U.~Beser},
  \textsc{H.\,I. Wang},  \textsc{D.~Beljonne},  \textsc{A.~Narita},
  \textsc{K.~M{\"u}llen},  \textsc{D.~Turchinovich},  and
  \textsc{M.~Bonn}\iffalse Role of {Edge} {Engineering} in {Photoconductivity}
  of {Graphene} {Nanoribbons}\fi,
 \jr{Journal of the American Chemical Society} \textbf{139}, 7982--7988 (2017).


\bibitem{schluenzen_cpp16}% article
 \textsc{N.~Schl{\"u}nzen} and  \textsc{M.~Bonitz}\iffalse {Nonequilibrium
  Green Functions Approach to Strongly Correlated Fermions in Lattice
  Systems}\fi,
 \jr{Contributions to Plasma Physics} \textbf{56}(1), 5--91 (2016).


\bibitem{schluenzen_prb16}% article
 \textsc{N.~Schl{\"u}nzen},  \textsc{S.~Hermanns},  \textsc{M.~Bonitz},  and
  \textsc{C.~Verdozzi}\iffalse Dynamics of strongly correlated
  fermions:\textit{Ab initio} results for two and three dimensions\fi,
 \jr{Phys. Rev. B} \textbf{93}(Jan), 035107 (2016).


\bibitem{schluenzen_prb17}% article
 \textsc{N.~Schl\"unzen},  \textsc{J.\,P. Joost},
  \textsc{F.~Heidrich-Meisner},  and  \textsc{M.~Bonitz}\iffalse
  {Nonequilibrium dynamics in the one-dimensional Fermi-Hubbard model:
  Comparison of the nonequilibrium Green-functions approach and the density
  matrix renormalization group method}\fi,
 \jr{Phys. Rev. B} \textbf{95}(Apr), 165139 (2017).


\bibitem{lipavski_prb_86}% article
 \textsc{P.~Lipavsk\'y},
  \textsc{V.~\ifmmode\,\check{S}\else\,\v{S}\fi{}pi\ifmmode\,\check{c}\else
  \v{c}\fi{}ka},  and  \textsc{B.~Velick\'y}\iffalse Generalized kadanoff-baym
  ansatz for deriving quantum transport equations\fi,
 \jr{Phys. Rev. B} \textbf{34}(Nov), 6933--6942 (1986).


\bibitem{hermanns_2014_hubbard}% article
 \textsc{S.~Hermanns},  \textsc{N.~Schlünzen},  and
  \textsc{M.~Bonitz}\iffalse Hubbard nanoclusters far from equilibrium\fi,
 \jr{Phys. Rev. B} \textbf{90}, 125111 (2014).


\bibitem{balzer_prb16}% article
 \textsc{K.~Balzer},  \textsc{N.~Schl{\"u}nzen},  and
  \textsc{M.~Bonitz}\iffalse Stopping dynamics of ions passing through
  correlated honeycomb clusters\fi,
 \jr{Phys. Rev. B} \textbf{94}(Dec), 245118 (2016).


\bibitem{balzer_prl_18}% article
 \textsc{K.~Balzer},  \textsc{M.~Rasmussen},  \textsc{N.~Schl{\"u}nzen},
  \textsc{J.\,P. Joost},  and  \textsc{M.~Bonitz}\iffalse Doublon formation by
  ions impacting a strongly correlated finite lattice system\fi,
 \jr{arXiv:1801.05267} (2018),
submitted for publication.


\bibitem{joost_pss_18}% article
 \textsc{J.\,P. Joost},  \textsc{N.~Schl{\"u}nzen},  and
  \textsc{M.~Bonitz}\iffalse Nonequilibrium electron dynamics in graphene
  nanoribbons - a nonequilibrium green function approach within an extended
  hubbard model\fi,
 \jr{physica status solidi (b)} (2018),
submitted for publication, this volume.


\bibitem{keldysh64}% article
 \textsc{L.~Keldysh}\iffalse Diagram technique for nonequilibrium processes\fi,
 \jr{Soviet Phys. JETP} \textbf{20}, 1018 (1965),
(Zh.~Eksp.~Teor.~Fiz.~\textbf{47}, 1515 (1964)).


\othercit
\bibitem{kadanoff-baym-book}% book
 \textsc{L.~Kadanoff} and  \textsc{G.~Baym},
Quantum Statistical Mechanics (Benjamin, New York, 1962).


\othercit
\bibitem{bonitz_qkt}% book
 \textsc{M.~Bonitz},
Quantum Kinetic Theory, 2 edition, Teubner-Texte zur Physik (Springer, 2016).


\bibitem{bonitz-etal.96pla}% article
 \textsc{M.~Bonitz} and  \textsc{D.~Kremp}\iffalse Kinetic energy relaxation
  and correlation time of nonequilibrium many-particle systems\fi,
 \jr{Physics Letters A} \textbf{212}(1–2), 83 -- 90 (1996).


\bibitem{bonitz-etal.96jpb}% article
 \textsc{M.~Bonitz},  \textsc{D.~Kremp},  \textsc{D.\,C. Scott},
  \textsc{R.~Binder},  \textsc{W.\,D. Kraeft},  and  \textsc{H.\,S.
  K\"ohler}\iffalse Numerical analysis of non-{M}arkovian effects in
  charge-carrier scattering: one-time versus two-time kinetic equations\fi,
 \jr{J. Phys.: Cond. Matt.} \textbf{8}(33), 6057 (1996).


\bibitem{bonitz96pla}% article
 \textsc{M.~Bonitz}\iffalse Correlation time approximation in non-markovian
  kinetics\fi,
 \jr{Physics Letters A} \textbf{221}(1–2), 85 -- 93 (1996).


\bibitem{kremp-etal.97ap}% article
 \textsc{D.~Kremp},  \textsc{M.~Bonitz},  \textsc{W.~Kraeft},  and
  \textsc{M.~Schlanges}\iffalse Non-{M}arkovian {B}oltzmann equation\fi,
 \jr{Annals of Physics} \textbf{258}(2), 320 -- 359 (1997).


\othercit
\bibitem{pngf1}% book
 \textsc{M.~Bonitz},  \textsc{R.~Nareyka},  and  \textsc{D.~Semkat},
{Progress in Nonequilibrium Green's Functions: Proceedings of the Conference
  "Kadanoff-Baym Equations: Progress and Perspectives for Many-body Physics",
  Rostock, Germany, 20-24 September 1999}, Progress in Nonequilibrium Green's
  Functions (World Scientific, 2000).


\othercit
\bibitem{pngf2}% book
 \textsc{M.~Bonitz} and  \textsc{D.~Semkat},
{Progress in Nonequilibrium Green's Functions II}, Progress in Nonequilibrium
  Green's Functions (World Scientific, 2003).


\bibitem{bonitz_pngf3}% article
 \textsc{M.~Bonitz} and  \textsc{A.~Filinov}\iffalse Progress in nonequilibrium
  green's functions iii\fi,
 \jr{Journal of Physics: Conference Series} \textbf{35}(1) (2006).


\bibitem{bonitz_jpcs_10}% article
 \textsc{M.~Bonitz} and  \textsc{K.~Balzer}\iffalse {Progress in Nonequilibrium
  Green's Functions IV}\fi,
 \jr{Journal of Physics: Conference Series} \textbf{220}(1), 011001 (2010).


\bibitem{vanleeuwen_jpcs_13}% article
 \textsc{R.~van Leeuwen},  \textsc{R.~Tuovinen},  and
  \textsc{M.~Bonitz}\iffalse {Progress in Nonequilibrium Green's Functions V
  (PNGF V)}\fi,
 \jr{Journal of Physics: Conference Series} \textbf{427}(1), 011001 (2013).


\bibitem{verdozzi_jpcs16}% article
 \textsc{C.~Verdozzi},  \textsc{A.~Wacker},  \textsc{C.\,O. Almbladh},  and
  \textsc{M.~Bonitz}\iffalse {Progress in Non-equilibrium Green's Functions
  (PNGF VI)}\fi,
 \jr{Journal of Physics: Conference Series} \textbf{696}(1), 011001 (2016).


\othercit
\bibitem{stefanucci_cambridge_2013}% book
 \textsc{G.~Stefanucci} and  \textsc{R.~van Leeuwen},
Nonequilibrium Many-Body Theory of Quantum Systems (Cambridge University Press,
  Cambridge, 2013).


\bibitem{bonitz_pss_18_keldysh}% article
 \textsc{M.~Bonitz},  \textsc{A.~Jauho},  \textsc{M.~Sadovskii},  and
  \textsc{S.~Tikhodeev}\iffalse In memoriam leonid keldysh\fi,
 \jr{physica status solidi (b)} (2018),
submitted for publication, this issue.


\bibitem{hermanns_psc_12}% article
 \textsc{S.~Hermanns},  \textsc{K.~Balzer},  and  \textsc{M.~Bonitz}\iffalse
  The non-equilibrium {Green function approach to inhomogeneous quantum
  many-body systems using the generalized Kadanoff–Baym ansatz}\fi,
 \jr{Physica Scripta} \textbf{2012}(T151), 014036 (2012).


\othercit
\bibitem{hermanns_phd_16}% phdthesis
 \textsc{S.~Hermanns},
Nonequilibrium Green functions. Selfenergy approximation techniques,,
PhD thesis, Kiel University, Kiel, FRG, 2016,
unpublished.


\othercit
\bibitem{balzer_2013_nonequilibrium}% book
 \textsc{K.~Balzer} and  \textsc{M.~Bonitz},
Nonequilibrium {Green}'s Functions Approach to Inhomogeneous Systems, No.~867
  in Lecture Notes in Physics (Springer, Berlin, Heidelberg, 2013).


\bibitem{von_friesen_2010_kadanoff}% article
 \textsc{M.\,P. von Friesen},  \textsc{C.~Verdozzi},  and  \textsc{C.\,O.
  Almbladh}\iffalse Kadanoff-{Baym} dynamics of {Hubbard} clusters: Performance
  of many-body schemes, correlation-induced damping and multiple steady and
  quasi-steady states\fi,
 \jr{Phys. Rev. B} \textbf{82}, 155108 (2010).


\bibitem{schluenzen_prb17_comment}% article
 \textsc{N.~Schl\"unzen},  \textsc{J.\,P. Joost},  and
  \textsc{M.~Bonitz}\iffalse {Comment on ``On the unphysical solutions of the
  Kadanoff-Baym equations in linear response: Correlation-induced homogeneous
  density-distribution and attractors''}\fi,
 \jr{Phys. Rev. B} \textbf{96}(Sep), 117101 (2017).


\bibitem{banyai_prl_95}% article
 \textsc{L.~B\'anyai},  \textsc{D.\,B.\,T. Thoai},  \textsc{E.~Reitsamer},
  \textsc{H.~Haug},  \textsc{D.~Steinbach},  \textsc{M.\,U. Wehner},
  \textsc{M.~Wegener},  \textsc{T.~Marschner},  and  \textsc{W.~Stolz}\iffalse
  Exciton--lo-phonon quantum kinetics: Evidence of memory effects in bulk
  gaas\fi,
 \jr{Phys. Rev. Lett.} \textbf{75}(Sep), 2188--2191 (1995).


\othercit
\bibitem{haug_2008_quantum}% book
 \textsc{H.~Haug} and  \textsc{A.\,P. Jauho},
Quantum Kinetics in Transport and Optics of Semiconductors (Springer, 2008).


\bibitem{lorke_jpcs_06}% article
 \textsc{M.~Lorke},  \textsc{T.\,R. Nielsen},  \textsc{J.~Seebeck},
  \textsc{P.~Gartner},  and  \textsc{F.~Jahnke}\iffalse Quantum kinetic effects
  in the optical absorption of semiconductor quantum-dot systems\fi,
 \jr{Journal of Physics: Conference Series} \textbf{35}(1), 182 (2006).


\bibitem{seebeck_prb_05}% article
 \textsc{J.~Seebeck},  \textsc{T.\,R. Nielsen},  \textsc{P.~Gartner},  and
  \textsc{F.~Jahnke}\iffalse Polarons in semiconductor quantum dots and their
  role in the quantum kinetics of carrier relaxation\fi,
 \jr{Phys. Rev. B} \textbf{71}(Mar), 125327 (2005).


\bibitem{kremp_99_pre}% article
 \textsc{D.~Kremp},  \textsc{T.~Bornath},  \textsc{M.~Bonitz},  and
  \textsc{M.~Schlanges}\iffalse Quantum kinetic theory of plasmas in strong
  laser fields\fi,
 \jr{Phys. Rev. E} \textbf{{\bf 60}}(Oct), 4725--4732 (1999).


\bibitem{bonitz_99_cpp}% article
 \textsc{M.~Bonitz},  \textsc{T.~Bornath},  \textsc{D.~Kremp},
  \textsc{M.~Schlanges},  and  \textsc{W.\,D. Kraeft}\iffalse Quantum kinetic
  theory for laser plasmas. dynamical screening in strong fields\fi,
 \jr{Contrib. Plasma Phys.} \textbf{{\bf 39}}(4), 329--347 (1999).


\bibitem{haberland_01_pre}% article
 \textsc{H.~Haberland},  \textsc{M.~Bonitz},  and  \textsc{D.~Kremp}\iffalse
  Harmonics generation in electron-ion collisions in a short laser pulse\fi,
 \jr{Phys. Rev. E} \textbf{{\bf 64}}(Jul), 026405 (2001).


\bibitem{kwong-etal.98pss}% article
 \textsc{N.~Kwong},  \textsc{M.~Bonitz},  \textsc{R.~Binder},  and
  \textsc{H.~K\"ohler}\iffalse Semiconductor {K}adanoff-{B}aym equations
  results for optically excited electron-hole plasmas semiconductor quantum
  wells\fi,
 \jr{phys. stat. sol. (b)} \textbf{{\bf 206}}, 197 (1998).


\bibitem{hermanns_jpcs13}% article
 \textsc{S.~Hermanns},  \textsc{K.~Balzer},  and  \textsc{M.~Bonitz}\iffalse
  Few-particle quantum dynamics–comparing nonequilibrium {G}reen functions
  with the generalized {K}adanoff–{B}aym ansatz to density operator
  theory\fi,
 \jr{Journal of Physics: Conference Series} \textbf{427}(1), 012008 (2013).


\bibitem{balzer_jpcs13}% article
 \textsc{K.~Balzer},  \textsc{S.~Hermanns},  and  \textsc{M.~Bonitz}\iffalse
  {The generalized Kadanoff-Baym ansatz. Computing nonlinear response
  properties of finite systems}\fi,
 \jr{J. Physics Conf. Ser.} \textbf{427}(1), 012006 (2013).


\bibitem{bonitz_cpp13}% article
 \textsc{M.~Bonitz},  \textsc{S.~Hermanns},  and  \textsc{K.~Balzer}\iffalse
  {Dynamics of Hubbard Nano-Clusters Following Strong Excitation}\fi,
 \jr{Contributions to Plasma Physics} \textbf{53}, 778--787 (2013).


\bibitem{bonitz-etal.99epjb}% article
 \textsc{M.~Bonitz},  \textsc{D.~Semkat},  and  \textsc{H.~Haug}\iffalse
  Non-{L}orentzian spectral functions for {C}oulomb quantum kinetics\fi,
 \jr{Europ. Phys. J. B} \textbf{9}, 309 (1999).


\bibitem{spicka_ijmpb_14}% article
 \textsc{V.~Spicka},  \textsc{B.~Velicky},  and  \textsc{A.~Kalvova}\iffalse
  Electron systems out of equilibrium: Nonequilibrium green's function
  approach\fi,
 \jr{International Journal of Modern Physics B} \textbf{28}(23), 1430013
  (2014).


\bibitem{kalvova_epl_18}% article
 \textsc{A.~Kalvova},  \textsc{B.~Velicky},  and  \textsc{V.~Spicka}\iffalse
  Generalized master equation for a molecular bridge improved by vertex
  correction to the generalized kadanoff-baym ansatz\fi,
 \jr{EPL (Europhysics Letters)} \textbf{121}(6), 67002 (2018).


\othercit
\bibitem{hopian_2018_phd}%
 \textsc{M.~Hopian},
Theoretical developments for the real-time description and control of nanoscale
  systems, 2018.


\bibitem{verdozzi_gkba-note}% article
 \textsc{M.~Hopian} and  \textsc{C.~Verdozzi}\iffalse Initial correlated states
  for the generalized kadanoff–baym ansatz without adiabatic switching-on of
  interactions in closed systems\fi,
 \jr{arXiv:1808.03520} (2018),
submitted for publication.


\bibitem{DANIELEWICZ_84_ap2}% article
 \textsc{P.~Danielewicz}\iffalse Quantum theory of nonequilibrium processes ii.
  application to nuclear collisions\fi,
 \jr{Annals of Physics} \textbf{152}(2), 305 -- 326 (1984).


\bibitem{semkat_00_jmp}% article
 \textsc{D.~Semkat},  \textsc{D.~Kremp},  and  \textsc{M.~Bonitz}\iffalse
  Kadanoff--{B}aym equations and non-{M}arkovian {B}oltzmann equation in
  generalized {T}-matrix approximation\fi,
 \jr{J. Math. Phys.} \textbf{{\bf 41}}(11), 7458--7467 (2000).


\othercit
\bibitem{kremp-springer}% book
 \textsc{D.~Kremp},  \textsc{M.~Schlanges},  and  \textsc{W.~Kraeft},
Quantum Statistics of Nonideal Plasmas (Springer, 2005).


\bibitem{karlsson_gkba18}% article
 \textsc{D.~Karlsson},  \textsc{R.~van Leeuwen},  \textsc{E.~Perfetto},  and
  \textsc{G.~Stefanucci}\iffalse The generalized kadanoff–baym ansatz with
  initial correlations\fi,
 \jr{arXiv:1806.05639} (2018),
submitted for publication.


\bibitem{semkat_cpp_03}% article
 \textsc{D.~Semkat},  \textsc{M.~Bonitz},  and  \textsc{D.~Kremp}\iffalse
  Relaxation of a quantum many-body system from a correlated initial state. a
  general and consistent approach\fi,
 \jr{Contributions to Plasma Physics} \textbf{43}(5-6), 321--325 (2003).


\bibitem{bonitz_jpcs_13}% article
 \textsc{M.~Bonitz},  \textsc{S.~Hermanns},  \textsc{K.~Kobusch},  and
  \textsc{K.~Balzer}\iffalse Nonequilibrium {G}reen function approach to the
  pair distribution function of quantum many-body systems out of
  equilibrium\fi,
 \jr{Journal of Physics: Conference Series} \textbf{427}(1), 012002 (2013).


\bibitem{eckstein-pes_2008}% article
 \textsc{M.~Eckstein} and  \textsc{M.~Kollar}\iffalse Measuring correlated
  electron dynamics with time-resolved photoemission spectroscopy\fi,
 \jr{Phys. Rev. B} \textbf{78}(Dec), 245113 (2008).


\bibitem{balzer_2016_stopping}% article
 \textsc{K.~Balzer},  \textsc{N.~Schlünzen},  and  \textsc{M.~Bonitz}\iffalse
  Stopping dynamics of ions passing through correlated honeycomb clusters\fi,
 \jr{Phys. Rev. B} \textbf{94}, 245118 (2016).


\bibitem{scharnke_jmp17}% article
 \textsc{M.~Scharnke},  \textsc{N.~Schl\"unzen},  and
  \textsc{M.~Bonitz}\iffalse Time reversal invariance of quantum kinetic
  equations: {Nonequilibrium Green functions formalism}\fi,
 \jr{Journal of Mathematical Physics} \textbf{58}(6), 061903 (2017).


\bibitem{bonitz_cpp18}% article
 \textsc{M.~Bonitz},  \textsc{M.~Scharnke},  and
  \textsc{N.~Schl\"unzen}\iffalse {Time‐reversal invariance of quantum
  kinetic equations II: Density operator formalism}\fi,
 \jr{Contributions to Plasma Physics} \textbf{58}(0) (2018).


\bibitem{landau_32}% article
 \textsc{L.\,D. Landau}\iffalse Zur theorie der energie\"ubertragung. ii\fi,
 \jr{Physikalische Zeitschrift der Sowjetunion} \textbf{2}, 46--51 (1932).


\bibitem{zener_32}% article
 \textsc{C.~Zener}\iffalse Non-adiabatic crossing of energy levels\fi,
 \jr{Proceedings of the Royal Society of London A: Mathematical, Physical and
  Engineering Sciences} \textbf{137}(833), 696--702 (1932).


\bibitem{kollar_prb_11}% article
 \textsc{M.~Kollar},  \textsc{F.\,A. Wolf},  and  \textsc{M.~Eckstein}\iffalse
  Generalized gibbs ensemble prediction of prethermalization plateaus and their
  relation to nonthermal steady states in integrable systems\fi,
 \jr{Phys. Rev. B} \textbf{84}(Aug), 054304 (2011).


\bibitem{joura_prb_15}% article
 \textsc{A.\,V. Joura},  \textsc{J.\,K. Freericks},  and  \textsc{A.\,I.
  Lichtenstein}\iffalse Long-lived nonequilibrium states in the hubbard model
  with an electric field\fi,
 \jr{Phys. Rev. B} \textbf{91}(Jun), 245153 (2015).


\bibitem{tokuno_pra_12}% article
 \textsc{A.~Tokuno},  \textsc{E.~Demler},  and  \textsc{T.~Giamarchi}\iffalse
  Doublon production rate in modulated optical lattices\fi,
 \jr{Phys. Rev. A} \textbf{85}(May), 053601 (2012).


\bibitem{pamperin_2015_many}% article
 \textsc{M.~Pamperin},  \textsc{F.\,X. Bronold},  and
  \textsc{H.~Fehske}\iffalse Many-body theory of the neutralization of
  strontium ions on gold surfaces\fi,
 \jr{Phys. Rev. B} \textbf{91}, 035440 (2015).


\bibitem{Moss_2009_Li+_AlCluster}% article
 \textsc{C.\,L. Moss},  \textsc{C.\,M. Isborn},  and  \textsc{X.~Li}\iffalse
  Ehrenfest dynamics with a time-dependent density-functional-theory
  calculation of lifetimes and resonant widths of charge-transfer states of
  ${\text{li}}^{+}$ near an aluminum cluster surface\fi,
 \jr{Phys. Rev. A} \textbf{80}, 024503 (2009).


\bibitem{Castro2012_H+Li4}% article
 \textsc{A.~Castro},  \textsc{M.~Isla},  \textsc{J.\,I. Mart\'inez},  and
  \textsc{J.\,A. Alonso}\iffalse Scattering of a proton with the li$_4$
  cluster: Non-adiabatic molecular dynamics description based on time-dependent
  density-functional theory\fi,
 \jr{Chemical Physics} \textbf{399}, 130 (2012).


\bibitem{Krasheninnikov2007}% article
 \textsc{A.\,V. Krasheninnikov},  \textsc{Y.~Miyamoto},  and
  \textsc{D.~Tom\'anek}\iffalse Role of electronic excitation in ion collisions
  with carbon nanostructures\fi,
 \jr{Phys. Rev. Lett.} \textbf{99}, 016104 (2007).


\bibitem{Bubin2012}% article
 \textsc{S.~Bubin},  \textsc{B.~Wang},  \textsc{S.~Pantelides},  and
  \textsc{K.~Varga}\iffalse Simulation of high-energy ion collicions with
  graphene fragments\fi,
 \jr{Phys. Rev. B} \textbf{85}, 235435 (2012).


\othercit
\bibitem{balzer-book}% book
 \textsc{K.~Balzer} and  \textsc{M.~Bonitz},
Nonequilibrium {G}reen's {F}unctions Approach to Inhomogeneous Systems
  (Springer, Berlin Heidelberg, 2013).


\bibitem{DORNHEIM_physrep18}% article
 \textsc{T.~Dornheim},  \textsc{S.~Groth},  and  \textsc{M.~Bonitz}\iffalse The
  uniform electron gas at warm dense matter conditions\fi,
 \jr{Physics Reports} \textbf{744}, 1 -- 86 (2018).


\bibitem{marini_2009_yambo}% article
 \textsc{A.~Marini},  \textsc{C.~Hogan},  \textsc{M.~Gr\"uning},  and
  \textsc{D.~Varsano}\iffalse yambo: An ab initio tool for excited state
  calculations\fi,
 \jr{Comp. Phys. Commun.} \textbf{180}, 1392 (2009).


\end{thebibliography}
%
% Replace the following example bibliography with your references
% before submission:
\providecommand{\WileyBibTextsc}{}
\let\textsc\WileyBibTextsc
\providecommand{\othercit}{}
\providecommand{\jr}[1]{#1}
\providecommand{\etal}{~et~al.}

% \begin{thebibliography}{[1]}

% \bibitem{bib1}%
%  F.\,M. Firstauthorfamilyname, F.\,M. Secondauthorfamilyname, and
%   C.~Lastauthorfamilyname,
%  Abbreviatedjournalname \textbf{volume}, page (year).

% \bibitem{bib2}%
%  F.~Examplename and  I.\,E. Anotherauthorname,
%  Phys. Status Solidi A \textbf{1}, 111 (2050).

% \bibitem{bib3}%
%  A.\,N.~Earlyview, E.\,X.~Ample, and Y.\,A.~Author,
%  Phys. Status Solidi A, DOI 10.1002/pssr.205001234  (2050).

% \bibitem{bib4}%
%  A.~Firstauthorname,  B.~Secondauthorname,  and
%   C.~Thirdauthorname,
% Here Goes the Title of the Book (Publisher, City, year), p.\,111.

% \bibitem{bib5}%
%  A.~Firsteditorname,  B.~Secondeditorname,  and
%   C.~Thirdeditorname (eds.),
% Here Goes the Title of the Edited Book (Wiley-VCH, Berlin, 2050), p.\,111.

% \bibitem{bib6}%
%  D.~Contributorname,
%  in: The Title of the Book, edited by The Name of the Editors, Followed by
%   the Title of the Series of Books (Publisher, City, year), chap.~1.

% \bibitem{bib7}%
%  A.~Lastbutnotleastname,
%  Proceedings 1st Dummy Conference on Citation Formatting, City,
%   Country, Part A (Publisher, City, year),  pp.\,1--11.

% \end{thebibliography}

\newpage

\section*{Graphical Table of Contents\\}
GTOC image:
\begin{figure}[h]%
\includegraphics[width=4cm,height=4cm]{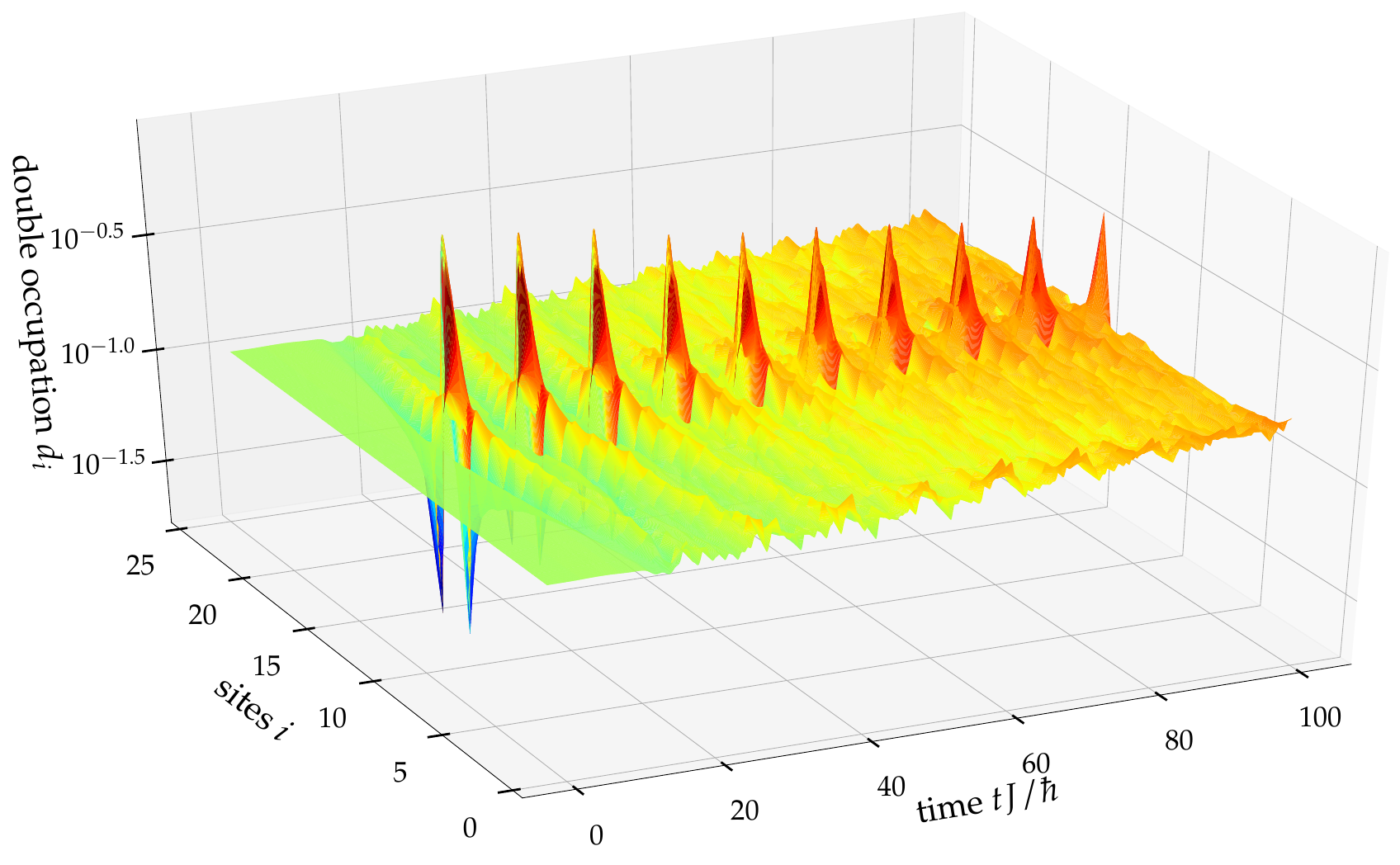}
\caption*{%
%  Your article will be published with a Graphical Abstract in the table of contents.
%  Please send a suggestion for an image (preferably full colour, size 4 cm x 4 cm). It may be specifically designed for the purpose, but should not show too many details or consist of several parts.
%Enclose a short descriptive and popular text on the general aim and value of your paper which may serve as an `appetizer' for the readers (40--70 words, not a figure caption, not the abstract text).
Excitation of correlated electron pairs (doublons) in a graphene nanoribbon by multiple ion impacts. Result of Nonequilibrium Green functions simulations.
}
\label{GTOC}
\end{figure}

\end{document}